\documentclass{article} 
\usepackage{iclr2027_conference,times}

\usepackage{amsmath,amsfonts,bm}

\def\eqref#1{equation~\ref{#1}}

\def\1{\bm{1}}

\DeclareMathAlphabet{\mathsfit}{\encodingdefault}{\sfdefault}{m}{sl}
\SetMathAlphabet{\mathsfit}{bold}{\encodingdefault}{\sfdefault}{bx}{n}

\usepackage{hyperref}
\usepackage{url}
\usepackage{array}
\usepackage[caption=false,font=normalsize,labelfont=sf,textfont=sf]{subfig}
\usepackage{textcomp}
\usepackage{stfloats}
\usepackage{verbatim}
\usepackage{graphicx}
\usepackage{amssymb}
\usepackage{mathtools}
\usepackage{amsthm}
\usepackage{booktabs}
\usepackage{multirow}
\usepackage[table,xcdraw]{xcolor}
\usepackage{color}
\usepackage{makecell}
\usepackage{arydshln}
\usepackage{enumitem}
\usepackage[capitalize,noabbrev]{cleveref}
\usepackage{cite}
\usepackage{pifont}
\usepackage{wasysym} 
\usepackage[normalem]{ulem}
\usepackage{algorithm}
\usepackage{algpseudocode}
\usepackage{caption}
\usepackage{wrapfig}
\usepackage{fontawesome5}
\newcommand{\std}[1]{\,{\LARGE\color{black!55}\ensuremath{\pm}#1}}
\newcommand{\metric}[2]{%
  #1\,{\color{black!70}\fontsize{0.90em}{1em}\selectfont / #2}%
}
\newcommand{\diff}[1]{\,{\LARGE\color{black!55}(#1)}}

\title{	Learning to Watermark Speech Synthesis Against Model-Driven Reconstruction }

\author{
Weizhi Liu\textsuperscript{1},
Yue Li\textsuperscript{2},
Hui Tian\textsuperscript{2},
Zhaoxia Yin\textsuperscript{1}\thanks{ Corresponding author}
\\
\textsuperscript{1}East China Normal University, Shanghai, China\\
\textsuperscript{2}Huaqiao University, Xiamen, China\\
}

\iclrfinalcopy 

\begin{document}
\lhead{}
\renewcommand{\headrulewidth}{0pt}

\maketitle

\vspace{-1.5em}
\begin{center}
    \href{https://lwzzz7.github.io/Thrive_Demo/}{
        \faGithub\hspace{0.4em}\textbf{Speech Demo Page}
    }
\end{center}

\begin{abstract}
Modern TTS systems increasingly generate synthetic speech at scale for diverse users.
This setting calls for content-level provenance that can verify the origin of released speech and attribute it to the requesting user, which generative watermarking can support by embedding multi-bit identifiers directly into synthesized speech.
Once released, however, speech may undergo heterogeneous learned transformations during distribution and editing, with reconstruction objectives that can preserve speech utility while affecting watermark recoverability differently.
We find that no single watermark carrier remains consistently reliable across reconstruction models, as its survival depends jointly on the embedded structure, reconstruction mechanism, and observation representation.
To this end, we propose \textsc{Thrive}, a multi-bit generative speech watermarking framework for modern autoregressive TTS, covering both discrete-token and continuous-representation generation under reconstruction attacks.
Specifically, \textsc{Rise} synchronizes watermark injection into intermediate representations with its continued integration into subsequent generation, while \textsc{Care} combines waveform and spectral experts using bit-wise reliability selection.
Experiments on both autoregressive paradigms show that \textsc{Thrive} preserves synthesis fidelity, achieves 87.6\% average recovery accuracy under reconstruction attacks, and supports source attribution over candidate sets of up to 10,000 identities.
\end{abstract}

\section{Introduction} 
Neural codec language models~\citep{cui2025SLM}, which autoregressively generate discrete codec tokens, have become a prominent paradigm for modern text-to-speech (TTS) synthesis, while recent systems extend to continuous acoustic representations~\citep{xie2025CTTS}.
As these models are deployed in TTS services, a single model may generate speech for many registered users.
Practical accountability therefore requires content-level provenance, including verifying the origin of released speech and attributing it to the requesting user~\citep{California2024LegInfo, AIGC2025CAC}.
Generative watermarking can support both objectives~\citep{zhao2025sok_wm} by embedding a specific multi-bit identifier directly into synthesized speech.
Once released, speech content may be transformed as it is distributed and edited across platforms.
Such transformations may occur in combination or in sequence in practice and are increasingly mediated by learned speech models, whose reconstruction objectives can preserve speech utility yet have markedly different effects on watermark recoverability.
Robustness tailored to a single reconstruction mechanism can therefore become brittle, as the watermark provider cannot control which sequence or combination of downstream transformations released speech will actually undergo.
When deliberately exploited to undermine watermark recovery while retaining usable speech, such learned transformations constitute \textbf{model-driven reconstruction attacks}.

Existing post-hoc methods embed watermarks into time-frequency or latent representations~\citep{san2024audioseal, chen2026latentmark}, while generative methods inject them during synthesis~\citep{liu2024groot, zhu2026law} or encode watermark behavior into model parameters~\citep{feng2025riwf, cheng2024hfgw}.
Despite these different integration points, existing research establishes the watermark through a fixed representation or model component and relies on the resulting acoustic structure to remain observable.
Heterogeneous reconstruction objectives can preserve, reorganize, or discard the same structure differently, leaving unresolved whether any fixed carrier can remain reliable across reconstruction attacks.

To test this assumption, we embed controlled spectro-temporal patterns into speech and trace their behavior across diverse acoustic and representation spaces.
We find that the same pattern can be preserved, reshaped, or suppressed differently across reconstruction models, while its observability varies across recovery representations. 
Watermark survival therefore depends jointly on the embedded structure, reconstruction mechanism, and observation representation, leaving no single carrier uniformly reliable across attacks.
This raises the question of \textit{how to maintain watermark recoverability as watermark behavior varies across heterogeneous reconstruction models}.

\begin{figure}[t]
    \centering
    \includegraphics[width=0.95\linewidth]{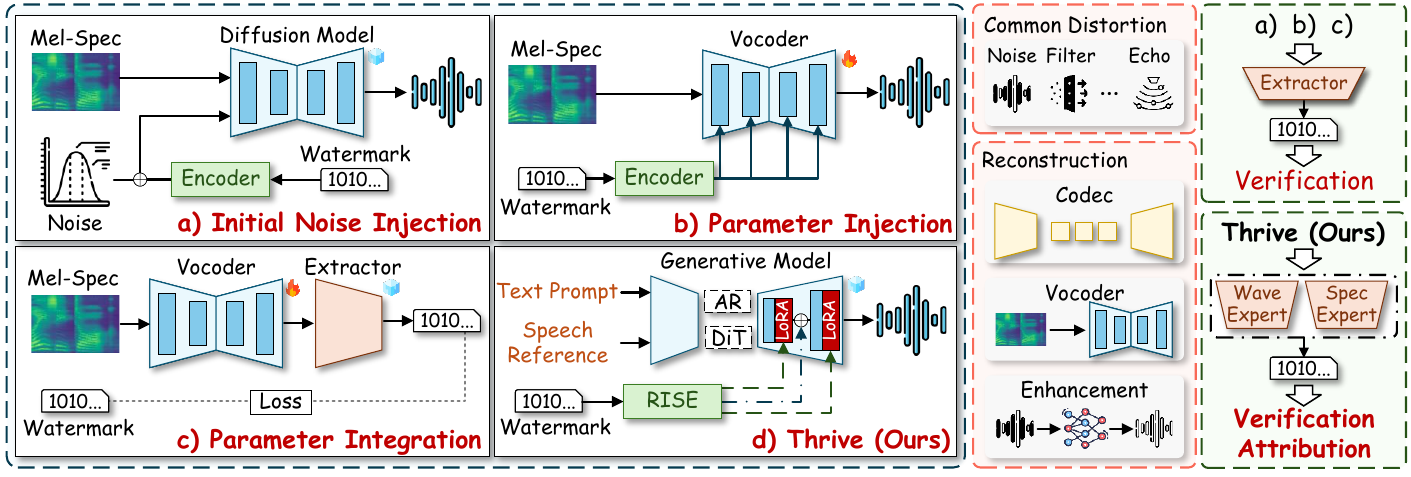}
    \caption{Comparison of generative speech watermarking paradigms. Existing approaches establish the watermark through a particular generation state, synthesis component, or model parameter, whereas \textsc{Thrive} coordinates watermark participation throughout generation and adapts recovery after model-driven reconstruction.}
    \label{fig_overview}
\end{figure}

To address this question, our key insight is to diversify watermark participation in generation and adapt its recovery after reconstruction.
We therefore propose \textsc{Thrive}, a generative speech watermarking framework for robust provenance verification and source attribution under model-driven reconstruction, as contrasted with existing paradigms in Fig.~\ref{fig_overview}.
\textsc{Rise} coordinates watermark injection into intermediate representations with its continued participation in subsequent generation, while \textsc{Care} combines waveform and spectral recovery through bit-wise reliability selection. 
We further decouple capability learning from reliability calibration so that the selector is trained against stable recovery behavior. 
Experiments on autoregressive TTS backbone with discrete tokens and continuous representations show that \textsc{Thrive} preserves synthesis fidelity while maintaining robust provenance verification under varied reconstruction and scalable multi-user source attribution.

Our contributions can be summarized as:
\begin{itemize}
    \item We study canonical reconstruction attacks within a unified benchmark and reveal that watermark survival depends jointly on the embedded structure, reconstruction mechanism, and recovery representation. Motivated by this finding, we develop \textsc{Thrive} for robust provenance verification and source attribution under heterogeneous reconstruction.

    \item \textsc{Thrive} integrates \textsc{Rise} for synchronized watermark participation across representation and generation, and \textsc{Care} for reliability-aware recovery across complementary waveform and spectral representations, enabling the framework to naturally accommodate continuous-representation autoregressive TTS.

    \item We validate \textsc{Thrive} on two modern autoregressive TTS architectures, demonstrating robust provenance verification and scalable multi-user source attribution under diverse reconstruction attacks while preserving generation fidelity.
\end{itemize}

\section{Related Work}
\textbf{Post-hoc speech watermarking} improves robustness by seeking embedding carriers that remain recoverable after signal processing or reconstruction.
WavMark~\citep{chen2023wavmark}, TBWM~\citep{liu2024timebre}, and True~\citep{li2026true} use frequency- or time-domain features, while AudioSeal~\citep{san2024audioseal}, XAttnMark~\citep{liu2025xattnmark}, and DiscreteWM~\citep{ji2025discretewm} extend watermark embedding to continuous or discrete latent representations.
FASW~\citep{li2026fasw} combines latent and frequency-domain features, whereas Latent-Mark~\citep{chen2026latentmark} constrains zero-bit watermark features to a latent subspace preserved by codec reconstruction.
These methods progressively explore different feature spaces while retaining the central strategy of finding carriers that remain stable after processing.

\textbf{Generative speech watermarking} has primarily studied multi-bit embedding in two-stage TTS pipelines, where the watermark is tied to a particular synthesis stage or generator component.
Groot~\citep{liu2024groot}, \citet{zhu2026law}, and \citet{rong2026anchorwm} inject the watermark at the initial stage of generation and rely on subsequent synthesis to propagate it to the output.
Parameter-based methods instead encode watermark functionality into the generator, with RIWF~\citep{feng2025riwf} modulating selected vocoder weights and HFGw~\citep{cheng2024hfgw} fine-tuning the vocoder to produce watermarked speech.
However, methods such as Groot and HFGw bind the embedded watermark to the trained model and require retraining to change the watermark.
Recent approaches extend watermarking to modern autoregressive models by steering discrete-token sampling, but \citet{wu2026alignis} and \citet{milis2026hipt} primarily support zero-bit detection.
In contrast, \textsc{Thrive} brings multi-bit generative watermarking to modern autoregressive TTS, including \textbf{continuous-representation generation} that has not been covered by prior autoregressive watermarking approaches.
It further supports utterance-specific identifiers \textbf{at inference time without retraining}, while explicitly addressing watermark robustness under heterogeneous model-driven reconstruction.

\section{Preliminaries}
\label{sec_prelimi}
\subsection{Problem definition}
\label{sec_pro_def}
Let $x_c$ denote the text prompt, $x_s$ the reference speech, and $\mathcal{G}$ a protected speech generative model serving multiple registered users.
Each registered user is assigned a distinct $l$-bit watermark $w \in \{0,1\}^{l}$.
Given $x_c$, $x_s$, and $w$, \textsc{Thrive} integrates $w$ into generation to produce the watermarked speech $s_w=\mathcal{G}(x_c,x_s,w)$.
The same model can embed user-specific watermarks at inference time without retraining.
After release, $s_w$ may undergo routine or adversarial processing, including signal distortions, lossy platform compression, and enhancement or restoration during cross-platform reuse.
We denote $\tilde{s}_w=\mathcal{P}(s_w)$, where $\mathcal{P}$ denotes such processing during distribution.
The recovery mechanism $\mathcal{D}$ then recovers:
\begin{equation}
    \hat{w} = \mathcal{D}(\tilde{s}_w)
    = \mathcal{D}\bigl(\mathcal{P}(\mathcal{G}(x_c,x_s,w))\bigr).
\end{equation}
Successful recovery requires $\hat{w}=w$ despite the applied processing.
The recovered watermark serves two purposes.
\ding{172} \textbf{Provenance Verification} links the processed speech $\tilde{s}_w$ to the protected model $\mathcal{G}$, 
while \ding{173} \textbf{Source Attribution} identifies the registered user who initiated the request.
Both objectives must be achieved while preserving \textbf{Generation Fidelity}.
Let $s$ denote the corresponding generation with watermark integration disabled.  
\textsc{Thrive} aims to preserve fidelity between $s_w$ and $s$ while reliably recovering $w$ after dissemination.

\subsection{Threat Model}
\textbf{Defender's abilities and goal.}
The defender deploys the protected model $\mathcal{G}$ as a speech generation service for registered users.
For each request, $\mathcal{G}$ embeds the watermark assigned to the requesting user and returns the watermarked speech $s_w$.
When $s_w$ later appears on public distribution channels, the defender uses $\mathcal{D}$ for provenance verification and source attribution.
The defender aims to preserve synthesis fidelity while maintaining reliable provenance verification and source attribution after dissemination and reconstruction.

\textbf{Attacker's abilities and goal.}
During dissemination, $s_w$ may undergo non-adversarial processing such as noise, filtering, resampling, and lossy compression, which we treat as normal conditions under which the watermark should remain recoverable.
By contrast, once $s_w$ becomes publicly accessible, an attacker may deliberately apply model-driven reconstruction, including neural codec regeneration, enhancement reconstruction, and vocoder resynthesis, to disrupt watermark recovery.
The attacker has access only to the released speech $s_w$ and may choose the reconstruction model and configuration, but has no access to the internal generation or recovery pipeline.
Denoting the selected reconstruction attack by $\mathcal{A}$, the attacker aims to make $\mathcal{D}(\mathcal{A}(s_w))\neq w$ while preserving speech content and perceptual quality.

\subsection{Analysis of Reconstruction Attacks}
\label{sec_analysis}
\begin{figure}[t]
    \centering
    \includegraphics[width=0.95\linewidth]{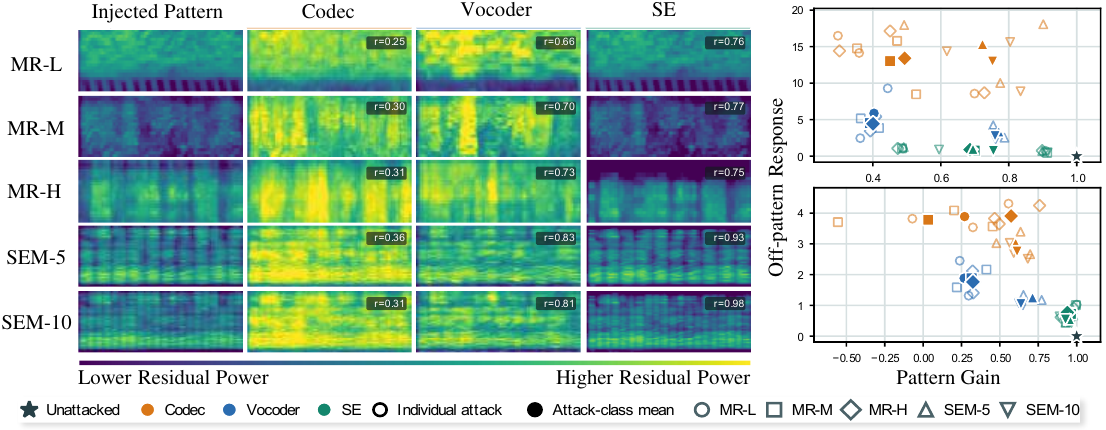}
    \caption{Controlled spectro-temporal pattern survival under reconstruction attacks. Mel residuals show reconstruction-dependent retention and deformation, while representation-space responses reveal that the same reconstructed pattern can appear differently across observation spaces.}
    \label{fig_reconstruction_analysis}
\end{figure}

To determine whether a stable watermark carrier can persist across reconstruction models, we track controlled spectro-temporal patterns at both acoustic and representation levels, following the ring-survival diagnostic for image editing~\citep{lu2025vine}.
For each reconstruction model, we compare paired speech with and without each pattern to isolate its post-reconstruction response.
As shown in the left panel of Fig.~\ref{fig_reconstruction_analysis}, temporal modulations are retained more strongly than localized spectral structures under enhancement reconstruction, while retention varies across codecs~\citep {guo2025codec_survey}, vocoders~\citep {o2026vocoder_review}, and enhancement reconstructions~\citep{o2024se_survey}.
The right panel shows that the same pattern-attack pair exhibits different retention and deformation in FACodec~\citep{Ju2024ns3} and WavLM~\citep{chen2022wavlm} representation spaces, indicating that watermark observability also depends on the recovery representation.
These findings show that watermark survival depends jointly on acoustic structure, reconstruction mechanism, and observation representation, so \textit{no single carrier or observation space can characterize reconstruction robustness.}
Further analysis is provided in Appendix~\ref{app:controlled_reconstruction}.

\section{Method}
\label{sec_method}
\subsection{Core Insight and Motivation}
\begin{figure}
    \centering
    \includegraphics[width=0.95\linewidth]{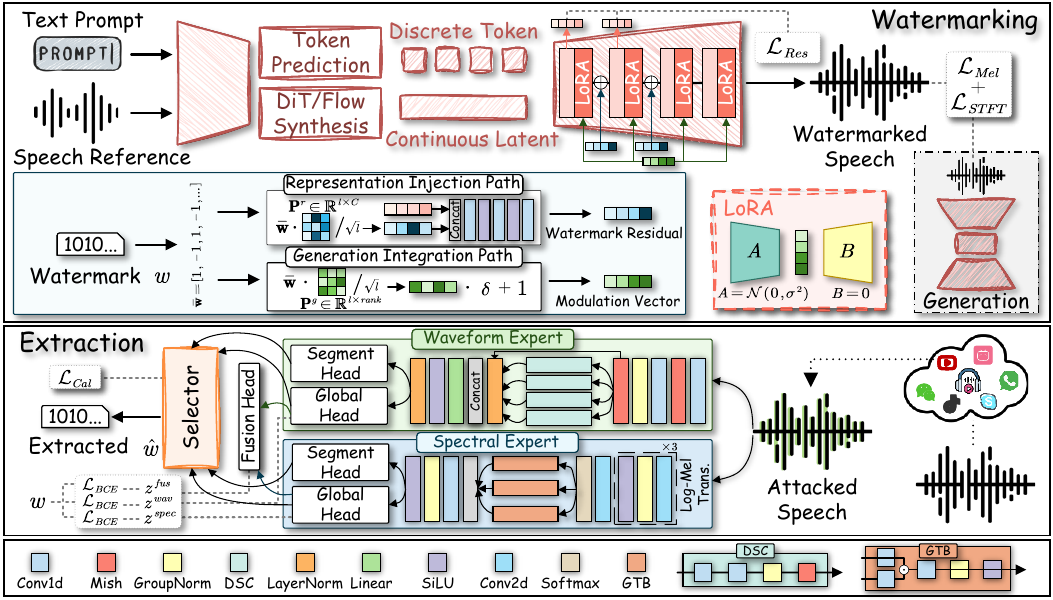}
    \caption{Overview of \textsc{Thrive}. \textsc{Rise} synchronizes watermark embedding across representation and generation by injecting a representation residual while modulating selected generation layers with the same message. \textsc{Care} recovers complementary waveform and spectral predictions and performs calibrated bit-wise selection for robust watermark recovery after reconstruction.}
    \label{fig_pipeline}
\end{figure}

Since reconstruction can alter both watermark structures and their observability, our core insight is to diversify how the watermark participates in generation and how it is recovered after reconstruction.
Rather than relying on a watermark injected at a single intermediate state to survive subsequent generation, \textsc{Rise} injects the message into evolving representations and conditions later generation on the same message, as illustrated in Fig.~\ref{fig_pipeline}.
Because reconstruction affects waveform and spectral representations differently, \textsc{Care} selects between their bit predictions according to estimated reliability.
We first train the embedding and recovery modules, then freeze them and calibrate only the selector so that its reliability estimates reflect fixed expert behavior.
Implementation details are provided in Appendix~\ref{app:method_details}.

\subsection{Representation-Generation Synchronized Embedding}
\label{sec_rise}
\textsc{Rise} synchronizes watermark injection with subsequent generation through a \textit{Representation Injection Path} and a \textit{Generation Integration Path}. 
The former writes the watermark into evolving intermediate representations, while the latter allows the same watermark to continue influencing subsequent generation. 
Given an $l$-bit watermark $\mathbf{w}\in\{0,1\}^{l}$, we first XOR it with a predefined key and map the result to the signed form $\bar{\mathbf{w}}\in\{-1,1\}^{l}$.

\textbf{Representation Injection Path.}
Let $\mathbf{h}_j\in\mathbb{R}^{T\times d}$ denote the intermediate speech representation at generation stage $j$.
We map the signed watermark into the representation space as $\mathbf{w}^{r}=\mathbf{P}^{r}\bar{\mathbf{w}}/\sqrt{l}$, where $\mathbf{P}^{r}\in\mathbb{R}^{d\times l}$ is a learnable projection matrix shared across the selected stages.
The projected watermark $\mathbf{w}^{r}$ is broadcast along the temporal dimension and content-adaptively modulated before residual injection:
\begin{equation}
\widetilde{\mathbf{h}}_j=\mathbf{h}_j+\Delta\mathbf{h}_j^{w},\qquad
\Delta\mathbf{h}_j^{w}=\lambda_j\left\{\mathbf{1}+\rho\tanh\!\left(\mathcal{F}_j\!\left[\operatorname{LayerNorm}(\mathbf{h}_j)\,\|\,\mathbf{w}^{r}\right]\right)\right\}\odot\mathbf{w}^{r},
\label{eq:rise_rep_injection}
\end{equation}
where $\Delta\mathbf{h}_j^{w}$ is the watermark residual and $\widetilde{\mathbf{h}}_j$ the resulting watermarked representation.
The fusion network $\mathcal{F}_j$ consists of two pointwise convolutions and a depthwise temporal convolution, with SiLU activations between them.
The coefficient $\lambda_j$ controls injection strength, while $\rho$ bounds the modulation within $[1-\rho,1+\rho]$.

\textbf{Generation Integration Path.}
To maintain watermark participation in subsequent generation, we introduce watermark-dependent rank-wise modulation into the low-rank branches of selected generation layers, following AquaLoRA~\citep{feng2024aqualora}. 
The signed watermark is mapped to a rank-modulation vector $\mathbf w^g=g(\mathbf 1+\delta\mathbf P^g\bar{\mathbf w}/\sqrt{l})$, where $\mathbf P^g\in\mathbb R^{r\times l}$ is a learnable projection, $\mathbf 1\in\mathbb R^r$ is the all-ones vector, $\delta$ controls the watermark-dependent modulation strength, and $g$ is a fixed modulation gain.
For a selected generation layer $m$, let $\mathbf h_m$ denote the intermediate representation received by that layer, which may already include watermark residuals injected at preceding representation stages. 
The modulated low-rank update is:
\begin{equation}
\widetilde{\mathcal W}_m(\mathbf h_m)=\mathcal W_m(\mathbf h_m)+\frac{\alpha}{r}\mathcal B_m\!\left[\mathbf w^g\odot\mathcal A_m(\mathbf h_m)\right],
\label{eq:generation_integration}
\end{equation}
where $\mathcal W_m$ denotes the frozen base transformation, $\mathcal A_m$ and $\mathcal B_m$ are trainable low-rank mappings, $r$ is the LoRA rank, and $\alpha$ is the fixed LoRA scaling factor shared across the selected layers. 
The modulation vector is shared across temporal positions. 
Because both $\mathbf w^r$ and $\mathbf w^g$ are derived from the input watermark, \textsc{Rise} supports different watermarks at inference time without retraining. 

\subsection{Cross-Representation Adaptive Recovery}
\label{sec_care}
Because reconstruction affects watermark recoverability differently across waveform and spectral representations, their relative reliability varies across attacks and bits.
The \textbf{Cross-Representation Adaptive Recovery} mechanism (\textsc{Care}) therefore uses complementary waveform and spectral experts and selects between their predictions for each bit according to estimated reliability.

\textbf{Complementary Recovery Experts.}
Given the processed speech $\widetilde{s}_w$, the Waveform Expert extracts watermark information directly from the waveform.
It extracts temporal features through a frontend and four dilated depthwise-separable convolutional (DSC) blocks, then fuses multi-level features with the deepest representation.
The Spectral Expert converts $\widetilde{s}_w$ into a normalized log-Mel spectrogram, extracts local time-frequency features with a convolutional encoder, aggregates the frequency dimension through learned attention, and applies parallel gated temporal branches (GTB).
Both experts use a Segment Head for temporal bit logits and a Global Head with attentive statistics pooling for utterance-level logits:
\begin{equation}
(\mathbf{z}^{\mathrm{wav}},\mathbf{U}^{\mathrm{wav}})=\mathcal{D}_{\mathrm{wav}}(\widetilde{s}_w),\qquad (\mathbf{z}^{\mathrm{spec}},\mathbf{U}^{\mathrm{spec}})=\mathcal{D}_{\mathrm{spec}}\!\left(\operatorname{LogMel}(\widetilde{s}_w)\right),
\label{eq:care_experts}
\end{equation}
where $\mathcal{D}_k$ comprises the corresponding encoder, Segment Head, and Global Head.
For $k\in\{\mathrm{wav},\mathrm{spec}\}$, $\mathbf{U}^{k}\in\mathbb{R}^{T_k\times l}$ contains segment-level logits, while $\mathbf{z}^{k}\in\mathbb{R}^{l}$ contains utterance-level logits.
The global logits serve as candidate predictions, while the segment-level logits capture temporal support for each bit.

Furthermore, a Fusion Head uses cross-attention between the aligned temporal representations to derive a spectral correction $\boldsymbol{\xi}\in\mathbb R^l$ to the waveform prediction:
\begin{equation}
\mathbf z^{\mathrm{fus}}=\mathbf z^{\mathrm{wav}}+\boldsymbol\gamma\odot\operatorname{clip}(\boldsymbol\xi,-\kappa,\kappa).
\label{eq:fusion}
\end{equation}
where $\boldsymbol\gamma\in[0,1]^l$ is estimated from both experts' prediction statistics and controls the bit-wise contribution of the spectral correction, while $\kappa$ bounds its magnitude. 
The fused logits $\mathbf z^{\mathrm{fus}}$ serve as the principal recovery output during capability learning, whereas the calibrated reliability selector later chooses directly between the two expert predictions.

\textbf{Bit-wise Reliability Selector.}
Because the relative reliability of the two experts can vary across bits, global confidence alone is insufficient for selecting between their predictions. 
For each expert-bit pair, \textsc{Care} estimates reliability from global confidence, global--local agreement, segment strength, and temporal consistency:
\begin{equation}
\begin{aligned}
\boldsymbol{\phi}_i^k=&\left[\tanh\!\left(\frac{|z_i^k|}{\tau}\right),\frac{1}{T_k}\sum_{t=1}^{T_k}\mathbb{I}\!\left[\operatorname{sgn}(U_{t,i}^k)=\operatorname{sgn}(z_i^k)\right],\right.\\[-2pt]
&\left.\operatorname*{median}_{t}\tanh\!\left(\frac{|U_{t,i}^k|}{\tau}\right),1-\operatorname{Std}_{t}\!\left(\tanh\!\left(\frac{U_{t,i}^k}{\tau}\right)\right)\right],
\end{aligned}
\label{eq:care_descriptor}
\end{equation}
where $\tau$ controls confidence scaling.
A shared scoring network maps each descriptor to a correctness score $q_i^k=\mathcal{S}(\boldsymbol{\phi}_i^k)+b_k$, with the expert-specific bias $b_k$ accounting for systematic confidence differences.
While $q_i^k$ captures bit-specific reliability, reconstruction may also favor one representation across the entire utterance.
We therefore define the bit-wise reliability difference $d_i=q_i^{\mathrm{spec}}-q_i^{\mathrm{wav}}$ and its utterance-level average $\bar{d}=l^{-1}\sum_{i=1}^{l}d_i$.
The final prediction is selected by:
\begin{equation}
\widehat{z}_i=\begin{cases}z_i^{\mathrm{spec}},&d_i+\lambda_g\bar{d}>0,\\z_i^{\mathrm{wav}},&d_i+\lambda_g\bar{d}\leq0,\end{cases}\qquad \widehat{w}_i=\mathbb{I}\!\left[\widehat{z}_i\geq0\right],
\label{eq:care_selection}
\end{equation}
where $\lambda_g$ controls the contribution of the utterance-level preference. 
Applying this rule to all $l$ bits yields the recovered watermark $\widehat{\mathbf{w}}$, with each bit selected according to expert-specific reliability and the utterance-level representation preference.

\subsection{Capability Learning and Reliability Calibration}
\label{sec_training}
Reliable expert selection requires established recovery capabilities that remain fixed during calibration. We therefore separate training into capability learning and reliability calibration.

\textbf{Capability Learning.}
We freeze the pretrained base generator and jointly optimize \textsc{Rise}, the Fusion Head, and both recovery experts.
Before watermark recovery, an attack simulator produces $\widetilde{s}_w=\mathcal{T}(s_w)$ using either the identity or a sampled training transformation, with details in Appendix~\ref{app:attack_settings}.
The recovery objective jointly supervises the fused and expert predictions:
\begin{equation}
\mathcal{L}_{\mathrm{wm}}=\operatorname{BCE}(\mathbf{z}^{\mathrm{fus}},\mathbf{w})+\lambda_{\mathrm{wav}}\operatorname{BCE}(\mathbf{z}^{\mathrm{wav}},\mathbf{w})+\lambda_{\mathrm{spec}}\operatorname{BCE}(\mathbf{z}^{\mathrm{spec}},\mathbf{w}).
\label{eq:recovery_loss}
\end{equation}

To preserve synthesis fidelity, $\mathcal L_{\mathrm{mel}}(s_w,s)$ and $\mathcal L_{\mathrm{stft}}(s_w,s)$ penalize Mel-spectrogram and multi-scale STFT differences between $s_w$ and its unwatermarked counterpart $s$. We further regularize the representation residuals as $\mathcal L_{\mathrm{res}}=\sum_{j\in\mathcal J}\operatorname{mean}[(\Delta\mathbf h_j^w)^2]$, where $\mathcal J$ denotes the selected injection stages and the mean is taken over all elements of each residual tensor. The overall objective is:
\begin{equation}
\mathcal L_{\mathrm{cap}}=\mathcal L_{\mathrm{wm}}+\lambda_{\mathrm{mel}}\mathcal L_{\mathrm{mel}}(s_w,s)+\lambda_{\mathrm{stft}}\mathcal L_{\mathrm{stft}}(s_w,s)+\lambda_{\mathrm{res}}\mathcal L_{\mathrm{res}}.
\label{eq:cap_loss}
\end{equation}

\textbf{Reliability Calibration.}
After capability learning, we freeze \textsc{Rise} and both recovery experts and calibrate only the shared scorer $\mathcal{S}$ and expert-specific biases $b_k$ on sampled processed speech.
For expert $k\in\{\mathrm{wav},\mathrm{spec}\}$, $c_i^k=\mathbb{I}[\mathbb{I}(z_i^k\geq0)=w_i]$ indicates whether bit $i$ is recovered correctly, and $q_i^k$ is treated as its correctness logit.
With $\lambda_g=\frac{1}{2}$, the selector in Eq.~(\ref{eq:care_selection}) chooses the spectral expert when $\Delta_i=d_i+\frac{1}{2}\bar d>0$ and the waveform expert otherwise.
For the ranking term, we use the disagreement set $\mathcal{I}=\{i\mid c_i^{\mathrm{wav}}\neq c_i^{\mathrm{spec}}\}$, where exactly one expert recovers the bit correctly:
\begin{equation}
\mathcal{L}_{\mathrm{cal}}=
\frac{1}{2}\sum_{k\in\{\mathrm{wav},\mathrm{spec}\}}
\operatorname{BCE}(\mathbf{q}^{k},\mathbf{c}^{k})
+
\frac{1}{2\max(1,|\mathcal{I}|)}
\sum_{i\in\mathcal{I}}
\log\!\left(1+\exp\!\left[-\Delta_i\left(c_i^{\mathrm{spec}}-c_i^{\mathrm{wav}}\right)\right]\right),
\label{eq:cal_loss}
\end{equation}
The first term calibrates each expert's score against bitwise correctness, while the second encourages the selector to prefer the correct expert on disagreement bits and vanishes when $\mathcal{I}$ is empty.

\begin{table*}[t]
  \centering
  \caption{Fidelity and watermark recovery performance. Values are mean $\pm$ standard deviation except FAD, which is computed at the corpus level. TPR is measured at 1\% FPR. Among generative watermarking methods, the best results are highlighted in \textbf{bold} and the second-best are \uline{underlined}. $\uparrow$ and $\downarrow$ indicate that higher and lower values are better, respectively.}
  \huge
  \resizebox{0.95\textwidth}{!}{
  \begin{tabular}{l*{10}{c}}
    \toprule
    \multirow{2}{*}{Method (bps)} & \multicolumn{4}{c}{Objective Fidelity} & \multicolumn{2}{c}{Perceptual Quality} & \multicolumn{2}{c}{FAD} & \multicolumn{2}{c}{Recovery} \\
    \cmidrule(lr){2-5}\cmidrule(lr){6-7}\cmidrule(lr){8-9}\cmidrule(lr){10-11}
    & STOI $\uparrow$ & PESQ $\uparrow$ & SSIM $\uparrow$ & MCD $\downarrow$ & DNSMOS $\uparrow$ & NISQA $\uparrow$ & VGGish $\downarrow$ & CLAP $\downarrow$ & ACC $\uparrow$ & TPR $\uparrow$ \\
    \midrule
    \multicolumn{11}{c}{LJSpeech (In-Distribution Dataset)} \\
    \midrule
    \multicolumn{11}{l}{\textit{Post-hoc Watermarking}} \\
    \midrule
    AudioSeal (16)
    & 0.998\std{0.001} & 4.425\std{0.073} & 0.978\std{0.007}
    & 0.818\std{0.160} & 3.202\std{0.211} & 3.875\std{0.625}
    & 0.077 & 0.004 & 1.000 & 1.000 \\

    WavMark (32)
    & 1.000\std{0.001} & 4.495\std{0.084} & 0.973\std{0.022}
    & 1.278\std{0.455} & 3.132\std{0.239} & 3.708\std{0.723}
    & 0.435 & 0.013 & 1.000 & 1.000 \\

    TBWM (100)
    & 0.986\std{0.007} & 3.929\std{0.214} & 0.937\std{0.013}
    & 1.539\std{0.138} & 3.213\std{0.208} & 3.678\std{0.621}
    & 0.855 & 0.019 & 1.000 & 1.000 \\

    \midrule
    \multicolumn{11}{l}{\textit{Generative Watermarking}} \\
    \midrule

    Groot (100)
    & \uline{0.958}\std{0.022} & 3.340\std{0.315} & 0.907\std{0.023}
    & 1.285\std{0.192} & \textbf{3.277}\std{0.214} & \uline{3.906}\std{0.624}
    & \textbf{0.058} & \textbf{0.002} & \uline{0.995} & 1.000 \\

    RIWF (16)
    & 0.957\std{0.018} & 2.414\std{0.317} & 0.814\std{0.047}
    & 6.310\std{1.503} & \uline{3.135}\std{0.315} & 3.098\std{0.651}
    & 3.501 & 0.102 & 0.993 & 1.000 \\

    HFGw (20)
    & 0.943\std{0.017} & 2.672\std{0.275} & \uline{0.935}\std{0.015}
    & \uline{0.907}\std{0.068} & 2.877\std{0.367} & 2.212\std{0.556}
    & 1.790 & 0.186 & 0.989 & 1.000 \\

    \midrule
    \rowcolor[HTML]{f7fcfd}
    Ours [Qwen3-TTS 0.6B] (16)
    & 0.958\std{0.073} & \uline{3.565}\std{0.768} & 0.850\std{0.087}
    & 1.754\std{0.136} & 2.638\std{0.429} & 2.807\std{1.165}
    & 1.355 & 0.086 & \textbf{0.999} & 1.000 \\

    \rowcolor[HTML]{f7fcfd}
    Ours [Dots.tts 2B] (16)
    & \textbf{0.999}\std{0.003} & \textbf{4.210}\std{0.259} & \textbf{0.980}\std{0.004}
    & \textbf{0.777}\std{0.103} & 3.066\std{0.317} & \textbf{4.094}\std{0.548}
    & \uline{0.604} & \uline{0.037} & \textbf{0.999} & 1.000 \\
    
    \midrule
    \multicolumn{11}{c}{LibriTTS (OOD Dataset)} \\
    \midrule
    \rowcolor[HTML]{f7fcfd}
    Ours [Qwen3-TTS] (16)
    & 0.969 \std{0.050} & 3.747\std{0.566} & 0.858\std{0.075}
    & 1.781\std{0.157} & 2.271\std{0.431} & 3.159\std{1.007}
    & 1.539 & 0.089 & 0.999 & 1.000 \\

    \rowcolor[HTML]{f7fcfd}
    Ours [Dots.tts] (16)
    & 0.955\std{0.197} & 4.045\std{0.479} & 0.980\std{0.004}
    & 0.739\std{0.135} & 2.617\std{0.286} & 3.827\std{0.812}
    & 0.412 & 0.052 & 0.999 & 0.999 \\
    
    \midrule
    \multicolumn{11}{c}{AISHELL-3 (OOD Dataset)} \\
    \midrule
    \rowcolor[HTML]{f7fcfd}
    Ours [Qwen3-TTS] (16)
    & 0.902\std{0.268} & 3.820\std{0.439} & 0.874\std{0.048}
    & 1.763\std{0.146} & 2.667\std{0.353} & 3.291\std{0.798}
    & 1.243 & 0.083 & 0.999 & 1.000 \\

    \rowcolor[HTML]{f7fcfd}
    Ours [Dots.tts] (16)
    & 0.887\std{0.314} & 4.069\std{0.471} & 0.979\std{0.004}
    & 0.764\std{0.138} & 2.600\std{0.325} & 3.399\std{1.045}
    & 0.551 & 0.067 & 0.999 & 1.000 \\
    \bottomrule
    \end{tabular}
}
  \label{tab_fidelity}
\end{table*}

\section{Experiments}
Following the objectives and the reconstruction analysis in Sec.~\ref{sec_prelimi}, we first describe the experimental setup (Sec.~\ref{sec_setup}) and then evaluate \textsc{Thrive} along four dimensions: generation fidelity and clean recovery (Sec.~\ref{sec_gen}), provenance verification (Sec.~\ref{sec_verify}), source attribution (Sec.~\ref{sec_attri}), and the contributions of \textsc{Rise} and \textsc{Care} (Sec.~\ref{sec_ablation}).

\subsection{Experimental Setup}
\label{sec_setup}
\textbf{Datasets and Baselines.}
Experiments use LJSpeech~\citep{Ito2017ljspeech} (single-speaker English), LibriTTS~\citep{zen2019libritts} (multi-speaker English), and AISHELL-3~\citep{shi2021aishell} (multi-speaker Chinese). We instantiate \textsc{Thrive} on Qwen3-TTS (0.6B)~\citep{hu2026qwen3} with discrete codec tokens and Dots.tts (2B)~\citep{lian2026dots} with continuous acoustic representations. Baselines include the post-hoc methods WavMark~\citep{chen2023wavmark} and AudioSeal~\citep{san2024audioseal}, Timbre Watermark~\citep{liu2024timebre}, and the generative methods Groot~\citep{liu2024groot}, RIWF~\citep{feng2025riwf}, and HFGw~\citep{cheng2024hfgw}.

\textbf{Evaluation Metrics.}
We report bit accuracy (ACC) for watermark recovery, STOI~\citep{taal2010stoi}, PESQ~\citep{recommendation2001pesq}, SSIM~\citep{wang2004ssim}, and MCD~\citep{kubichek1993mel} for paired speech fidelity, DNSMOS~\citep{reddy2022dnsmos} and NISQA~\citep{mittag21nisqa} for non-intrusive perceptual quality, and FAD for distribution-level generation quality.

\textbf{Implementation Details.}
We optimize trainable parameters with AdamW. During capability learning, \textsc{Rise}, the Fusion Head, and both recovery experts are trained for 80 epochs with batch size 2 and learning rates of $1\times10^{-4}$ for the low-rank generation path and $5\times10^{-4}$ for other modules. We use a three-epoch warm-up, cosine annealing after epoch 60, and introduce sampled processing from epoch 20. The reliability selector is then calibrated for 10 epochs at $1\times10^{-4}$ with all other modules frozen. Training uses a single NVIDIA RTX 4090 GPU.

\begin{table*}[t]
\centering
\caption{Provenance verification under common distortions and reconstruction attacks on LJSpeech. Entries are ACC / TPR@1\%FPR and Ours-Q/D denote Qwen3-TTS/Dots.tts.}
\LARGE
\setlength{\tabcolsep}{10pt}

\resizebox{0.95\textwidth}{!}{
\begin{tabular}{lcccccc>{\columncolor[HTML]{f7fcfd}}c>{\columncolor[HTML]{f7fcfd}}c}
\toprule
& \multicolumn{3}{c}{\textit{Post-hoc Watermarking}}
& \multicolumn{5}{c}{\textit{Generative Watermarking}} \\
\cmidrule(lr){2-4}
\cmidrule(lr){5-9}

& AudioSeal (16) & WavMark (32) & TBWM (100) & Groot (100) & RIWF (16) & HFGw (20) 
& \cellcolor{white}\textbf{Ours-Q (16)} & \cellcolor{white}\textbf{Ours-D (16)} \\
\midrule

Clean & \metric{0.999}{1.000} & \metric{1.000}{1.000} & \metric{1.000}{1.000} & \metric{0.995}{1.000} & \metric{0.993}{1.000} & \metric{0.989}{1.000} & \metric{0.999}{1.000} & \metric{0.999}{1.000} \\

\midrule
\noalign{\vskip -2pt}
\multicolumn{9}{l}{\textit{Signal Distortion}} \\
\noalign{\vskip -2pt}
\midrule

Gau. Noi. 
& \metric{0.808}{0.380} & \metric{0.561}{0.022} & \metric{0.725}{0.986} & \metric{\textbf{0.995}}{1.000} & \metric{0.951}{0.959} & \metric{0.926}{0.987} & \metric{0.983}{0.998} & \metric{\uline{0.993}}{1.000} \\
Low-pass 
& \metric{0.978}{0.995} & \metric{1.000}{1.000} & \metric{0.993}{1.000} & \metric{\uline{0.985}}{1.000} & \metric{0.525}{0.004} & \metric{0.914}{0.976} & \metric{\textbf{0.999}}{1.000} & \metric{\textbf{0.999}}{1.000} \\
Echo 
& \metric{0.969}{0.987} & \metric{0.859}{0.971} & \metric{0.947}{1.000} & \metric{\textbf{0.986}}{1.000} & \metric{0.593}{0.016} & \metric{\uline{0.983}}{1.000} & \metric{0.958}{0.973} & \metric{0.962}{0.979} \\
Resample 
& \metric{1.000}{1.000} & \metric{0.998}{1.000} & \metric{0.900}{1.000} & \metric{\uline{0.992}}{1.000} & \metric{0.741}{0.176} & \metric{0.799}{0.625} & \metric{\textbf{0.999}}{1.000} & \metric{\textbf{0.999}}{1.000} \\
Quant. 
& \metric{0.995}{1.000} & \metric{0.951}{1.000} & \metric{0.947}{1.000} & \metric{0.996}{1.000} & \metric{\uline{0.990}}{0.999} & \metric{0.986}{1.000} & \metric{\textbf{0.999}}{1.000} & \metric{\textbf{0.999}}{1.000} \\

\midrule
\noalign{\vskip -2pt}
\multicolumn{9}{l}{\textit{Traditional Compression}} \\
\noalign{\vskip -2pt}
\midrule

MP3 
& \metric{0.964}{0.982} & \metric{0.740}{0.543} & \metric{0.749}{0.997} & \metric{\uline{0.992}}{1.000} & \metric{0.965}{0.982} & \metric{0.845}{0.812} & \metric{\textbf{0.999}}{1.000} & \metric{\textbf{0.999}}{1.000} \\
Opus 
& \metric{0.455}{0.001} & \metric{0.904}{0.997} & \metric{0.861}{1.000} & \metric{0.760}{0.999} & \metric{0.614}{0.024} & \metric{\uline{0.835}}{0.774} & \metric{\textbf{0.999}}{1.000} & \metric{\textbf{0.999}}{1.000} \\

\midrule
\noalign{\vskip -2pt}
\multicolumn{9}{l}{\textit{Neural Codec Regeneration}} \\
\noalign{\vskip -2pt}
\midrule

EnCodec 
& \metric{0.509}{0.003} & \metric{0.496}{0.003} & \metric{0.582}{0.190} & \metric{\textbf{0.973}}{1.000} & \metric{0.803}{0.363} & \metric{0.578}{0.033} & \metric{0.891}{0.748} & \metric{\uline{0.967}}{0.986} \\
FACodec 
& \metric{0.507}{0.003} & \metric{0.523}{0.007} & \metric{0.553}{0.073} & \metric{0.749}{0.997} & \metric{0.634}{0.034} & \metric{0.581}{0.035} & \metric{\uline{0.788}}{0.309} & \metric{\textbf{0.842}}{0.524} \\
STCodec 
& \metric{0.505}{0.002} & \metric{0.502}{0.004} & \metric{0.553}{0.073} & \metric{\textbf{0.949}}{1.000} & \metric{\uline{0.853}}{0.577} & \metric{0.595}{0.046} & \metric{0.762}{0.230} & \metric{0.818}{0.422} \\
HiFiCodec 
& \metric{0.508}{0.003} & \metric{0.498}{0.003} & \metric{0.544}{0.051} & \metric{0.724}{0.984} & \metric{0.680}{0.073} & \metric{0.547}{0.018} & \metric{\uline{0.854}}{0.580} & \metric{\textbf{0.943}}{0.942} \\
VoiceCraft 
& \metric{0.505}{0.002} & \metric{0.490}{0.002} & \metric{0.551}{0.068} & \metric{\textbf{0.980}}{1.000} & \metric{0.605}{0.020} & \metric{0.480}{0.004} & \metric{0.796}{0.338} & \metric{\uline{0.837}}{0.504} \\

\midrule
\noalign{\vskip -2pt}
\multicolumn{9}{l}{\textit{Vocoder Resynthesis}} \\
\noalign{\vskip -2pt}
\midrule

WaveGlow 
& \metric{0.511}{0.003} & \metric{0.508}{0.005} & \metric{0.783}{1.000} & \metric{0.775}{1.000} & \metric{0.549}{0.007} & \metric{0.443}{0.001} & \metric{\textbf{0.933}}{0.914} & \metric{\uline{0.892}}{0.753} \\
HiFi-GAN 
& \metric{0.506}{0.002} & \metric{0.509}{0.005} & \metric{0.816}{1.000} & \metric{0.725}{0.986} & \metric{0.548}{0.006} & \metric{0.544}{0.017} & \metric{\textbf{0.860}}{0.606} & \metric{\uline{0.769}}{0.248} \\
BigVGAN 
& \metric{0.520}{0.003} & \metric{0.506}{0.004} & \metric{0.884}{1.000} & \metric{0.727}{0.987} & \metric{0.642}{0.039} & \metric{0.497}{0.005} & \metric{\textbf{0.899}}{0.785} & \metric{\uline{0.882}}{0.709} \\
Vocos 
& \metric{0.509}{0.003} & \metric{0.531}{0.009} & \metric{0.938}{1.000} & \metric{0.715}{0.975} & \metric{0.645}{0.041} & \metric{0.640}{0.101} & \metric{\textbf{0.881}}{0.703} & \metric{\uline{0.863}}{0.619} \\
DiffWave 
& \metric{0.503}{0.002} & \metric{0.491}{0.003} & \metric{0.690}{0.917} & \metric{0.743}{0.995} & \metric{0.533}{0.005} & \metric{0.487}{0.004} & \metric{\textbf{0.925}}{0.886} & \metric{\uline{0.868}}{0.646} \\
PriorGrad 
& \metric{0.507}{0.002} & \metric{0.490}{0.003} & \metric{0.688}{0.911} & \metric{0.740}{0.994} & \metric{0.532}{0.004} & \metric{0.489}{0.004} & \metric{\textbf{0.926}}{0.886} & \metric{\uline{0.870}}{0.646} \\
WaveFM 
& \metric{0.527}{0.004} & \metric{0.469}{0.001} & \metric{0.929}{1.000} & \metric{0.687}{0.906} & \metric{0.755}{0.210} & \metric{0.590}{0.043} & \metric{\uline{0.920}}{0.868} & \metric{\textbf{0.965}}{0.983} \\

\midrule
\noalign{\vskip -2pt}
\multicolumn{9}{l}{\textit{Enhancement Reconstruction}} \\
\noalign{\vskip -2pt}
\midrule

FRCRN 
& \metric{0.998}{1.000} & \metric{0.999}{1.000} & \metric{0.989}{1.000} & \metric{\textbf{0.996}}{1.000} & \metric{0.589}{0.015} & \metric{\uline{0.962}}{0.999} & \metric{0.902}{0.798} & \metric{0.805}{0.370} \\
Demucs 
& \metric{1.000}{1.000} & \metric{1.000}{1.000} & \metric{0.993}{1.000} & \metric{\textbf{0.995}}{1.000} & \metric{0.561}{0.008} & \metric{0.958}{0.999} & \metric{\uline{0.976}}{0.994} & \metric{0.895}{0.769} \\

\bottomrule
\end{tabular}
}
\label{tab_robustness}
\end{table*}

\subsection{Generation Fidelity and Clean Recovery}
\label{sec_gen}
We first examine \textit{whether \textsc{Thrive} can preserve speech fidelity while enabling reliable watermark recovery. }Table~\ref{tab_fidelity} highlights three key findings.
\ding{172} Both implementations achieve near-perfect clean recovery, with an ACC of 0.999 and a TPR of 1.000 at 1\% FPR. Dots.tts retains particularly high fidelity (PESQ 4.210 and SSIM 0.980), while Qwen3-TTS preserves intelligibility (STOI 0.958) but exhibits larger acoustic deviation. 
\ding{173} This reliable recovery extends beyond the training distribution. On LibriTTS and AISHELL-3, both backbones retain near-perfect watermark recovery despite changes in speaker diversity and language. 
\ding{174} The approach applies consistently across distinct autoregressive representations. Qwen3-TTS generates discrete codec tokens, whereas Dots.tts operates on continuous acoustic representations, showing that \textsc{Thrive} supports reliable multi-bit recovery while preserving generation fidelity in both settings.

\subsection{Provenance verification}
\label{sec_verify}
To \textit{assess provenance verification after dissemination and reconstruction}, we evaluate \textsc{Thrive} under common distortions and model-driven reconstruction attacks. Table~\ref{tab_robustness} highlights three key findings.
\ding{172} \textsc{Thrive} remains highly stable under routine processing and robust across diverse reconstruction attacks. Dots.tts performs better under neural codec regeneration, whereas Qwen3-TTS is stronger under most vocoder resynthesis models and both enhancement models.
\ding{173} Robustness varies substantially across reconstruction mechanisms, with clear performance crossings among \textsc{Thrive} backbones and competing methods. This echoes the reconstruction-dependent carrier survival identified in Sec.~\ref{sec_analysis}, showing that robustness depends on the interaction between watermarking and reconstruction rather than a universally stable carrier.
\ding{174} Despite this heterogeneity, \textsc{Thrive} maintains reliable verification across all three reconstruction families, preserving provenance after dissemination and model-driven reconstruction.

\subsection{Source Attribution}
\label{sec_attri}
\begin{wrapfigure}{r}{0.6\textwidth}
\vspace{-0.8em}
\centering
\includegraphics[width=0.98\linewidth]{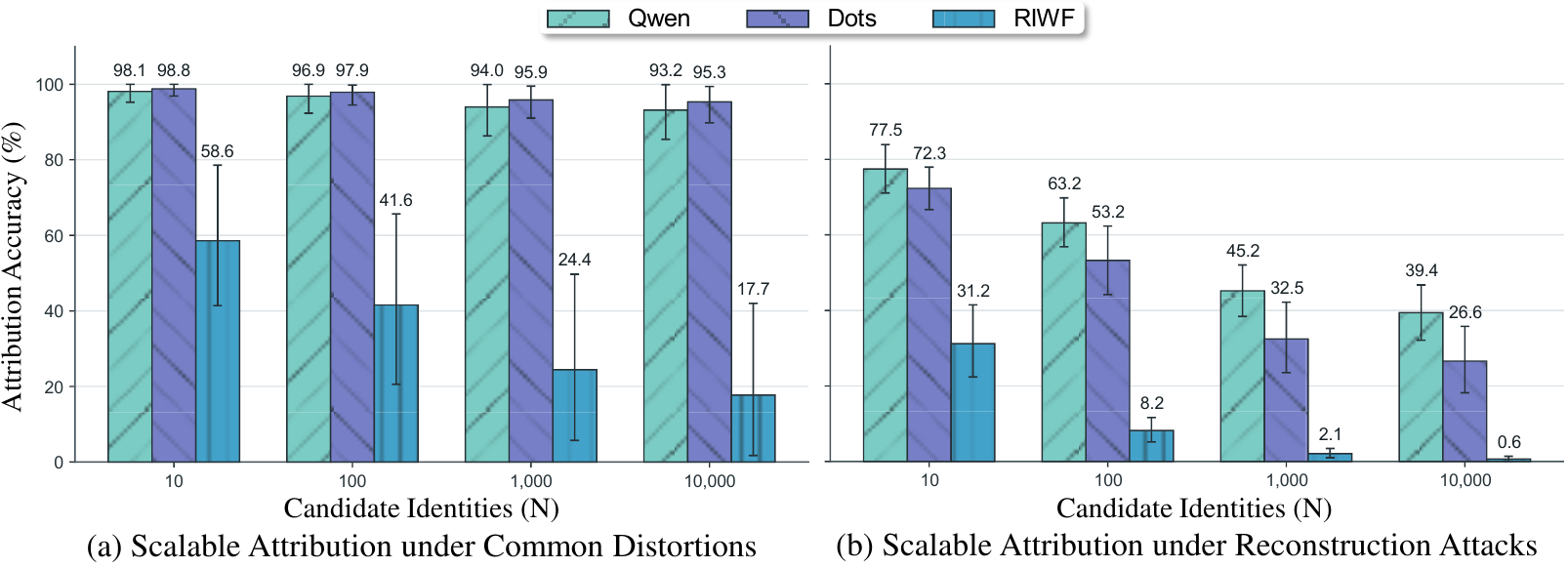}
\caption{Source attribution across increasing candidate-set sizes. Results are averaged over common distortions in (a) and across reconstruction attacks in (b).}
\label{fig_attribution}
\vspace{-0.2em}
\end{wrapfigure}

Following XAttnMark~\citep{liu2025xattnmark}, we evaluate source attribution by nearest-Hamming retrieval over candidate pools from $10$ to $10{,}000$ under common processing and model-driven reconstruction, as shown in Fig.~\ref{fig_attribution}.
\ding{172} \textsc{Thrive} maintains strong attribution as the candidate population scales. Under common processing, Qwen3-TTS and Dots.tts retain 93.2\% and 95.3\% accuracy at $N=10{,}000$, whereas RIWF drops to 17.7\%.
\ding{173} Reconstruction makes attribution substantially harder, yet both \textsc{Thrive} variants consistently outperform RIWF across all candidate-set sizes. 
At $N=10{,}000$, Qwen3-TTS and Dots.tts retain 39.4\% and 26.6\% accuracy, compared with 0.6\% for RIWF. 
This difference further reflects reconstruction-dependent robustness.
\ding{174} These results show that the recovered watermark remains discriminative beyond provenance verification, supporting scalable multi-user attribution under both routine and reconstruction processing.

\subsection{Contributions of \textsc{Thrive}}
\label{sec_ablation}
\begin{wraptable}{r}{0.55\textwidth}
\vspace{-0.8em}
\centering
\caption{Ablation of \textsc{Thrive} on Qwen3-TTS. Parentheses indicate ACC changes relative to the full model.}

\Huge
\setlength{\tabcolsep}{2.5pt}
\renewcommand{\arraystretch}{1.08}
\resizebox{\linewidth}{!}{%
\begin{tabular}{lccccc}
\toprule
\multirow{2}{*}{Variant}
& \multicolumn{2}{c}{Quality}
& \multicolumn{3}{c}{Recovery ACC} \\
\cmidrule(lr){2-3}\cmidrule(lr){4-6}
& PESQ $\uparrow$
& NISQA $\uparrow$
& Clean $\uparrow$
& Codec Avg. $\uparrow$
& Vocoder Avg. $\uparrow$ \\
\midrule
\rowcolor[HTML]{F7FCFD}
\textbf{Full \textsc{Thrive}}
& 3.565 & 2.807
& 0.999 & 0.818 & 0.906 \\

\midrule
\multicolumn{6}{l}{\textit{Embedding Ablation}} \\
Generation only
& 3.755 & 3.313
& 0.999\diff{0.000}
& 0.675\diff{-0.143}
& 0.674\diff{-0.232} \\

Representation only
& 4.643 & 3.477
& 0.489\diff{-0.510}
& -- & -- \\

Static LoRA
& 4.156 & 3.451
& 0.489\diff{-0.510}
& -- & -- \\

\midrule
\multicolumn{6}{l}{\textit{Recovery Ablation}} \\

Waveform only
& 3.814 & 3.276
& 1.000\diff{+0.001}
& 0.657\diff{-0.161}
& 0.771\diff{-0.135} \\

Spectral only
& 3.408 & 3.315
& 0.938\diff{-0.061}
& 0.873\diff{+0.055}
& 0.771\diff{-0.135} \\

Without Selector
& 3.565 & 2.807
& 0.999\diff{0.000}
& 0.780\diff{-0.038}
& 0.844\diff{-0.062} \\
\bottomrule
\end{tabular}
}
\label{tab_ablation}
\vspace{-0.5em}
\end{wraptable}

Table~\ref{tab_ablation} ablates the embedding and recovery components of \textsc{Thrive} and highlights three key findings.
\ding{172} On the embedding side, generation-only preserves clean recovery but reduces codec and vocoder ACC by 0.143 and 0.232, respectively, whereas representation-only and static LoRA remain near chance-level on clean speech despite high synthesis quality. This shows that representation injection strengthens robustness when synchronized with continued watermark participation during generation.
\ding{173} On the recovery side, the waveform expert performs best on clean speech, whereas the spectral expert is substantially stronger under codec regeneration. Removing the selector further reduces codec and vocoder ACC by 0.038 and 0.062 while leaving clean recovery unchanged, supporting the contribution of reliability-based selection. These complementary recovery profiles echo the representation-dependent observability identified in Sec.~\ref{sec_analysis} and motivate adaptive recovery across representations.
\ding{174} Overall, the ablations support coordinating watermark participation across representations and generation while avoiding reliance on a single recovery representation under diverse reconstruction.

\section{Conclusion}
We investigate generative speech watermarking under model-driven reconstruction and show that watermark survival depends jointly on the embedded structure, reconstruction mechanism, and recovery representation.
Motivated by this finding, we propose \textsc{Thrive}, where \textsc{Rise} coordinates watermark participation across representation and generation, while \textsc{Care} performs reliability-aware cross-representation recovery.
Experiments on modern autoregressive TTS, including the continuous-representation setting, demonstrate high synthesis fidelity, robust recovery under heterogeneous reconstruction attacks and common distortions, and scalable source attribution.

\subsection*{AI use statement}
In this work, we used generative AI tools for sentence translation, manuscript polishing, and detailed checking of consistency throughout the paper.
We have not used generative AI tools to fabricate experiments or invent experimental results.
We have reviewed all AI-assisted work and verified the final manuscript content and reported results for correctness.
We take responsibility for the final content of this work, including text, claims, or artifacts produced with the aid of generative AI.



\bibliography{iclr2027_conference}
\bibliographystyle{iclr2027_conference}

\appendix
\section*{Appendix}

\section{Detailed Analysis of Reconstruction Attacks}
\label{app:controlled_reconstruction}

\subsection{Motivation and Analysis Scope}

Released speech may be regenerated by neural codecs, resynthesized by vocoders, or processed by speech enhancement models during distribution and reuse.
The watermark provider cannot determine in advance which reconstruction mechanism or specific model will be applied downstream, so robustness tailored to a single reconstruction path may not generalize to others.
This motivates a more basic question of whether any spectro-temporal structure can remain consistently preserved across these reconstruction mechanisms and whether the surviving information remains equally observable in a fixed representation.

We study this question using controlled spectro-temporal patterns whose acoustic structure is known before reconstruction.
By comparing the same pattern across reconstruction models and observation spaces, we can directly characterize what is preserved, how it is reshaped, and where the remaining response is observable.
Our analysis covers neural codec regeneration, vocoder resynthesis, and enhancement reconstruction, with multiple models in each family where possible to capture both cross-family and within-family variation.
The results show that neither a single acoustic pattern nor a single observation representation remains consistently reliable across reconstruction models.

\subsection{Controlled Spectro-Temporal Patterns}

We choose five controlled patterns along two interpretable acoustic axes, spectral location and temporal modulation.
The goal is not to prescribe a particular watermark carrier, but to test whether common structural properties exhibit consistent preservation under reconstruction.

\paragraph{Spectral patterns.}
We use three Mel-localized ripple patterns to examine whether reconstruction preferentially preserves particular spectral regions.
Speech is represented by an 80-bin power Mel spectrogram computed from 24-kHz audio using a 1,024-point Hann window and a hop length of 256.
The low, middle, and high patterns target Mel bins 0--19, 30--49, and 60--79 and are denoted MR-L, MR-M, and MR-H, respectively.
Within the selected region, we add a deterministic zero-mean ripple whose phase varies jointly across time and frequency.
The modified Mel spectrum is mapped back to a linear-frequency magnitude and reconstructed with 32 Griffin--Lim iterations.

\paragraph{Temporal patterns.}
We further use two speech-envelope modulations to examine temporal structure independently of nominal spectral location.
For modulation frequency $f\in\{5,10\}$~Hz, the unnormalized pattern is:
\begin{equation}
\widetilde{\delta}_{f}(s;t)=s(t)\sin(2\pi ft),
\label{eq:envelope_modulation}
\end{equation}
where multiplication by the speech waveform concentrates the modification in speech-active regions.
The two patterns are denoted SEM-5 and SEM-10.

To make comparisons reflect pattern structure rather than perturbation energy, every realized waveform pattern is normalized to 2\% of the RMS of its corresponding clean utterance,
\begin{equation}
\delta_m(s)=
\frac{\widetilde{\delta}_m(s)}
{\operatorname{RMS}(\widetilde{\delta}_m(s))+\epsilon}
\operatorname{RMS}(s)\cdot0.02.
\label{eq:controlled_rms}
\end{equation}
The same pattern realization and relative amplitude are used across reconstruction conditions.

\subsection{Reconstruction Models and Observation Spaces}

We evaluate seven reconstruction models spanning the three attack families considered in the main paper.
Neural codec regeneration includes EnCodec at 6~kbps, FACodec, and SpeechTokenizer.
Vocoder resynthesis includes DiffWave and HiFi-GAN.
Enhancement reconstruction includes FRCRN and Demucs.
Using multiple models within each family is important because a shared application category does not imply a shared reconstruction mechanism, and one objective of the analysis is to determine whether models within the same family preserve the same acoustic structures.

Each reconstructed pattern is examined in three observation spaces that expose complementary aspects of its survival.
The Mel space preserves explicit time--frequency organization and therefore shows whether the original acoustic structure remains directly visible.
The FACodec latent space provides a learned acoustic representation and reveals whether reconstruction retains or redistributes pattern information beyond its explicit Mel structure.
WavLM provides a complementary self-supervised speech representation and allows us to test whether the same reconstructed pattern appears differently under another learned representation.
Together, these spaces distinguish acoustic preservation from representation-dependent observability.

FACodec is used both as one reconstruction model and as a fixed observation encoder for all reconstruction conditions.
We therefore interpret the FACodec-on-FACodec case together with the Mel and WavLM results rather than treating it as an independent cross-architecture comparison.

\subsection{Measuring Pattern Survival}

We use 20 one-second LJSpeech segments sampled at 24~kHz.
Crossing 20 utterances, seven reconstruction models, and five pattern configurations yields 700 paired reconstruction observations.

For each utterance, we send both the original speech and its pattern-modified counterpart through the same reconstruction model.
Let $A$ denote the reconstruction operation and $\Psi$ an observation mapping.
The pattern responses before and after reconstruction are:
\begin{equation}
\Delta_m^{\Psi}(s)=\Psi(s+\delta_m(s))-\Psi(s),
\qquad
\Delta_{A,m}^{\Psi}(s)=\Psi(A(s+\delta_m(s)))-\Psi(A(s)).
\label{eq:paired_pattern_response}
\end{equation}
The second difference isolates the incremental response associated with the controlled pattern after both branches undergo the same reconstruction.
For stochastic models, the two branches use matched random seeds, and all features are compared over their common temporal extent.

We characterize the relationship between the pre-reconstruction response $\mathbf u=\operatorname{vec}(\Delta_m^{\Psi})$ and the post-reconstruction response $\mathbf v=\operatorname{vec}(\Delta_{A,m}^{\Psi})$ from three complementary perspectives.
Directional alignment measures whether the original pattern direction is preserved.
Pattern gain measures how much response remains along that direction.
Off-pattern response measures how much reconstruction redirects the pattern into other feature directions.
They are defined as:
\begin{align}
\operatorname{Align}(\mathbf u,\mathbf v)
&=
\frac{\langle\mathbf v,\mathbf u\rangle}
{\|\mathbf v\|_2\|\mathbf u\|_2+\epsilon},
\\
\operatorname{Gain}(\mathbf u,\mathbf v)
&=
\frac{\langle\mathbf v,\mathbf u\rangle}
{\|\mathbf u\|_2^2+\epsilon},
\\
\operatorname{Off}(\mathbf u,\mathbf v)
&=
\frac{\|\mathbf v-\operatorname{Gain}(\mathbf u,\mathbf v)\mathbf u\|_2}
{\|\mathbf u\|_2+\epsilon}.
\end{align}
Together, these quantities describe preservation, retained strength, and deformation rather than reducing reconstruction behavior to a single attenuation value.

Model-level results first average the paired measurements over utterances.
Class-level summaries then assign equal weight to reconstruction models within each family.
Fig.~\ref{fig_reconstruction_analysis} visualizes both the acoustic responses and their behavior in learned observation spaces.

\subsection{Analysis Results}

\paragraph{No spectro-temporal pattern is consistently preserved across reconstruction models.}
The relative advantage of a pattern changes with the reconstruction mechanism.
For Mel directional alignment, MR-L/MR-M/MR-H average 0.25/0.30/0.31 under neural codecs, approximately 0.66--0.73 under vocoders, and 0.75--0.77 under enhancement reconstruction.
SEM-5/SEM-10 reach 0.36/0.31, 0.83/0.81, and 0.93/0.98 in the same three families.
Temporal modulation is therefore substantially more stable under the tested vocoders and enhancement models, but this advantage does not persist consistently under neural codec regeneration.
The three localized spectral patterns likewise show no stable low-to-high ordering across codecs.
These results rule out a single spectral region or temporal structure as a consistently preserved carrier across reconstruction models.


\paragraph{Reconstruction behavior varies within and across model families.}
Models assigned to the same reconstruction family can preserve the same pattern very differently.
DiffWave is nearly uniform across the five patterns, with Mel directional alignment between 0.648 and 0.691, whereas HiFi-GAN strongly favors the two temporal patterns, whose alignments reach 0.971 and 0.938 compared with 0.651--0.771 for the localized spectral patterns.
The contrast is even stronger for enhancement reconstruction.
FRCRN reduces the three localized-pattern alignments to approximately 0.51--0.55 while retaining SEM-5 and SEM-10 at 0.873 and 0.968, whereas Demucs remains close to the unattacked direction for all five patterns.
The neural codecs also exhibit different pattern orderings.
Pattern survival is therefore governed by the specific reconstruction mechanism rather than by its coarse attack family alone.

\paragraph{Pattern observability changes across representations.}
Reconstruction not only changes how much of a pattern remains, but also changes the representation in which that response is most visible.
In FACodec space, neural codec regeneration retains nonzero projections of the temporal patterns while introducing substantial off-pattern components, indicating that part of the original structure is preserved but redistributed into different feature directions.
Representation dependence is also visible for vocoder resynthesis.
DiffWave appears nearly uniform across patterns in Mel space, yet in FACodec space its SEM-5/SEM-10 gains reach 0.754/0.758 compared with approximately 0.36--0.44 for its three localized spectral patterns.
Similarly, FRCRN substantially weakens localized Mel responses while leaving coherent responses observable in WavLM space.
The same reconstructed pattern can therefore appear attenuated, preserved, or redistributed depending on the observation representation.

\paragraph{Summary.}
The controlled analysis reveals that pattern survival is jointly determined by the embedded acoustic structure, the reconstruction mechanism, and the observation representation.
No single spectro-temporal pattern remains consistently preserved across reconstruction models, and no single observation space fully characterizes the information that survives reconstruction.
These findings motivate the two complementary design choices in \textsc{Thrive}, namely maintaining watermark participation throughout generation with \textsc{Rise} and recovering from complementary representations with \textsc{Care}.

\section{\textsc{Thrive}: Detailed Architecture and Implementation}
\label{app:method_details}

This section provides the architecture and implementation details omitted from the main text and follows the notation in Sec.~\ref{sec_method}.
The pretrained speech backbones remain frozen, while only the watermark-specific modules described below are optimized.

\subsection{Architecture Overview}

Figure~\ref{fig_pipeline} summarizes the generation and recovery flows of \textsc{Thrive}.
On the generation side, \textsc{Rise} uses the same input watermark through two synchronized paths.
The Representation Injection Path writes a watermark residual into selected intermediate representations, while the Generation Integration Path maintains message-dependent modulation in subsequent generation layers.
This design is instantiated at different decoder locations for Qwen3-TTS and Dots.tts without changing their original input--output interfaces.

On the recovery side, the processed speech $\widetilde{s}_w$ is sent to both the Waveform Expert and the Spectral Expert in \textsc{Care}.
Each expert produces utterance-level and segment-level bit logits.
During capability learning, the Fusion Head combines the two expert representations and provides the principal fused prediction, while both expert predictions are also supervised.
After capability learning, the experts are frozen and the reliability Selector is calibrated from their fixed recovery behavior.
At inference, the Fusion Head is not used for the final decision, and the calibrated Selector chooses directly between the waveform and spectral predictions for each bit.

\begin{table}[t]
\centering
\caption{Backbone-specific placement of the RISE paths.}
\small
\setlength{\tabcolsep}{4pt}
\begin{tabular}{p{0.17\linewidth}p{0.36\linewidth}p{0.36\linewidth}}
\toprule
Backbone & Representation Injection Path & Generation Integration Path \\
\midrule
Qwen3-TTS &
Outputs of \texttt{post\_upsample\_1} and
\texttt{post\_upsample\_2}; the feature residual is added at each named
stage. &
Full-decoder LoRA updates on 57 modules: pre\_conv.conv (Conv1d), the
q/k/v/o attention projections (Linear), and the gate/up/down MLP projections
(Linear). \\

Dots.tts &
Outputs of \texttt{decoder.ups[0]} and
\texttt{decoder.ups[1]}; the feature residual is added before the
corresponding \texttt{decoder.resblocks} AMP aggregation. &
LoRA updates on 48 modules: post\_proj (Conv1d), dec\_mi\_layer.1.lstm
(LSTM), decoder.conv\_pre (Conv1d), decoder.ups[0] and decoder.ups[1]
(ConvTranspose1d), and selected decoder.resblocks convolutions (Conv1d). \\
\bottomrule
\end{tabular}
\label{tab_rise_backbones}
\end{table}

\subsection{\textsc{Rise} Architecture and Embedding Flow}

RISE is attached only to the synthesis decoder of each backbone. The original text and reference-speech processing remain unchanged, and the same backbone produces either watermarked or unwatermarked speech depending on whether watermark integration is enabled. The Representation Injection Path and Generation Integration Path follow Eqs.~(2) and~(3), respectively, and are conditioned on the same user-specific watermark, while the original backbone parameters remain frozen.

Before carrier projection, the watermark is XOR-whitened with a secret-key stream and mapped to signed values. Two learnable carrier projections, initialized with orthogonal bases, map the keyed message into the decoder feature space and rank space for the two RISE paths, respectively.

For the Representation Injection Path, the projected watermark is broadcast along the temporal axis and content-adaptively modulated by $F_j$ following Eq.~(\ref{eq:rise_rep_injection}). We set the injection strengths $\lambda_j$ to $10^{-3}$ and $5\times10^{-4}$ at the two selected decoder stages, respectively. Each $F_j$ consists of two pointwise convolutions and a depthwise temporal convolution with SiLU activations, and the modulation is bounded by $\rho$ as defined in Eq.~(\ref{eq:rise_rep_injection}). Only valid non-padding positions are updated.

For Qwen3-TTS, representation residuals are injected at the outputs of \texttt{post\_upsample\_1} and \texttt{post\_upsample\_2}. For Dots.tts, they are injected at the outputs of \texttt{decoder.ups[0]} and \texttt{decoder.ups[1]}, before the corresponding \texttt{decoder.resblocks} AMP aggregation.

For the Generation Integration Path, both backbones use LoRA rank $r=64$ and scaling $\alpha=64$. The watermark-dependent modulation coefficient is $\delta=0.9$, and the fixed modulation gain is $g=2.0$. Let $r_m$ denote the signed keyed message projected into the rank space; the resulting modulation is $s_m=g(1+\delta r_m)=2.0(1+0.9r_m)$, shared across temporal positions. The backbone-specific LoRA placements are listed in Table~\ref{tab_rise_backbones}. The carrier projections and low-rank adapter parameters are trainable, while the original decoder weights remain frozen.

\subsection{\textsc{Care} Architecture and Recovery Flow}

\textsc{Care} uses complementary waveform and spectral recovery paths with a common hidden width of 256.
Both experts produce global logits $\mathbf z^k$ and segment-level logits $\mathbf U^k$, following Eq.~(\ref{eq:care_experts}).
The global logits provide candidate bit predictions, while the segment-level logits provide temporal evidence for reliability estimation.

\paragraph{Waveform Expert.}
The Waveform Expert operates directly on $\widetilde{s}_w$.
A strided one-dimensional frontend is followed by four depthwise-separable convolution blocks with dilation rates $1$, $2$, $4$, and $8$.
Features from multiple levels are projected to a common width and fused before the Segment Head and Global Head.
The Segment Head produces temporal bit logits, while the Global Head uses attentive statistics pooling to produce utterance-level logits.

\paragraph{Spectral Expert.}
The Spectral Expert converts $\widetilde{s}_w$ into an 80-bin log-Mel spectrogram at 24~kHz using a 1,024-point FFT and window and a hop length of 256.
A two-dimensional convolutional frontend extracts local time--frequency features, followed by learned frequency attention.
Three temporal branches with kernel sizes $3$, $7$, and $15$ model complementary temporal scales, with two gated temporal blocks in each branch.
The fused temporal representation is passed to Segment and Global Heads with the same output format as the waveform path.

\paragraph{Fusion and Reliability Selection.}
During capability learning, the Fusion Head applies four-head cross-attention between the aligned temporal representations, using waveform features as queries and spectral features as keys and values.
The resulting spectral correction is combined with the waveform prediction according to Eq.~(\ref{eq:fusion}).
The Fusion Head is used only during capability learning and does not determine the final inference output.

After the two experts are frozen, the Selector estimates their bit-wise reliability from the four descriptors in Eq.~(\ref{eq:care_descriptor}), namely global confidence, global--local sign agreement, median segment strength, and temporal stability.
A lightweight shared MLP maps these descriptors to expert correctness scores, with an expert-specific bias accounting for systematic confidence differences.
The final decision follows Eq.~(\ref{eq:care_selection}), combining the bit-wise score difference with the utterance-level representation preference.
No ground-truth watermark, attack identity, or clean reference is required at inference.

\subsection{Training and Inference Procedure}

\paragraph{Capability Learning.}
We first freeze the pretrained speech backbone and jointly optimize \textsc{Rise}, the Fusion Head, and both \textsc{Care} experts.
Capability learning runs for 80 epochs with batch size 2.
The low-rank generation path is optimized with learning rate $1\times10^{-4}$, while the remaining trainable modules use $5\times10^{-4}$.
We use a three-epoch warm-up and cosine annealing after epoch 60.

Epochs 1--19 use clean watermarked speech for recovery training.
From epoch 20 onward, an attack simulator produces the processed speech $\widetilde{s}_w=\mathcal{T}(s_w)$ before watermark recovery.
The training distribution assigns 85\% probability to reconstruction-oriented transformations and 15\% to broad signal augmentations.
The attack operators and their parameter ranges are provided in Appendix~\ref{app:attack_settings}.
The recovery predictions are supervised by Eq.~(\ref{eq:recovery_loss}), while synthesis fidelity and representation residual magnitude are controlled by Eq.~(\ref{eq:cap_loss}).

\paragraph{Reliability Calibration.}
After capability learning, \textsc{Rise}, the Fusion Head, and both recovery experts are frozen.
We calibrate only the shared reliability scorer and expert-specific biases for 10 additional epochs with learning rate $1\times10^{-4}$.
For each expert and bit, the correctness target is computed from the detached expert prediction and the ground-truth watermark.
The calibration objective in Eq.~(\ref{eq:cal_loss}) aligns reliability scores with expert correctness and trains the Selector to prefer the correct expert when their bit predictions disagree.
These correctness targets are used only during calibration.

\paragraph{Inference.}
For watermark generation, the text prompt and reference speech are processed by the protected TTS backbone while \textsc{Rise} conditions the synthesis decoder on the user-specific watermark.
Changing the watermark does not require retraining the backbone or watermark modules.

For recovery, the received waveform is processed independently by the Waveform and Spectral Experts.
The Selector computes the reliability descriptors from their global and segment-level predictions and selects one expert logit for each bit according to Eq.~(\ref{eq:care_selection}).
The selected logits are thresholded to obtain the keyed bit sequence, after which the inverse XOR operation recovers the original watermark.
The recovery procedure requires neither the applied attack nor an unprocessed reference signal.

\section{Attack Simulator and Robustness Settings}
\label{app:attack_settings}

\subsection{Attack Simulator}
\label{app:attack_simulator}

The Attack Simulator is applied to watermarked speech before recovery.
Capability learning uses clean watermarked speech during epochs 1--19.
From epoch 20 onward, one transformation is sampled for each mini-batch, with 85\% probability from the reconstruction-proxy branch and 15\% from the conventional augmentation branch.
The same sampling distribution is used during reliability calibration, where only the reliability selector is updated.
For non-differentiable operations, the transformed waveform is used in the forward pass with a straight-through gradient during backpropagation.

\paragraph{Conventional augmentations.}
Several conventional signal-processing operations are adapted from the distortion pipeline of Timbre Watermark~\citep{liu2024timebre}, with parameters adjusted to our 24-kHz training pipeline.
We additionally include codec compression and Mel-reconstruction variants as listed below.

\textbf{Mel-spectrogram reconstruction.}
Following the reconstruction procedure used in Timbre Watermark~\citep{liu2024timebre}, the peak-normalized waveform is converted to an 80-bin Mel representation, mapped back to a linear-frequency magnitude using the pseudoinverse Mel matrix, and reconstructed with 32 Griffin--Lim iterations.
We use three variants based on linear Mel magnitude, log-Mel magnitude, and log-Mel magnitude smoothed with a $3\times5$ filter.

\textbf{EnCodec.}
We perform an EnCodec encode--decode round trip at 6~kbps.

\textbf{Opus.}
We perform an Opus encode--decode round trip at 16~kbps.

\textbf{Gaussian noise.}
Additive white Gaussian noise is applied with an SNR uniformly sampled from 15--20~dB.

\textbf{Low-pass filtering.}
The cutoff frequency is uniformly sampled from 2--4~kHz.

\textbf{Amplitude scaling.}
The waveform amplitude is scaled by a factor sampled from $[0.90,1.10]$.

\textbf{Requantization.}
The waveform is uniformly requantized to 256 levels.

\textbf{Dither.}
TPDF dither is applied without noise shaping.

\textbf{Echo.}
A single delayed reflection is added with delay and gain independently sampled from $[0.1,0.5]$~s and $[0.1,0.5]$, respectively.

\textbf{Sample suppression.}
A contiguous 10--30\% portion of the waveform is zeroed at the beginning, middle, or end while preserving the waveform length.

\textbf{Time shift.}
The waveform is shifted left or right by 5--50~ms with equal probability and zero-padded at the exposed boundary.

\textbf{Resampling.}
The 24-kHz waveform is resampled to 16, 22.05, or 32~kHz and then converted back to 24~kHz.

\paragraph{Reconstruction proxies.}
To expose the recovery network to reconstruction behavior beyond conventional signal processing, we introduce five model-free transformations that approximate spectral smoothing, quantization, remapping, and selective suppression produced by learned reconstruction systems.

\textbf{Spectral-smoothing resynthesis.}
The STFT magnitude is smoothed with a $5\times5$ time--frequency average filter and attenuated toward high frequencies as
$0.45M+0.55M_{\mathrm{sm}}(1-0.35f^{1.7})$.
The waveform is reconstructed by inverse STFT using the original phase.

\textbf{Quantized-codec proxy.}
The peak-normalized waveform is quantized with a resolution uniformly sampled from 32--128 levels.
With probability $0.55$, a stride-one average filter with kernel size sampled from $\{3,4,5,6,7\}$ is additionally applied.

\textbf{Signed spectral remapping.}
The normalized frequency axis is divided into $[0,0.22)$, $[0.22,0.58)$, and $[0.58,1]$, with gains sampled from $[0.75,1.20]$, $[0.55,1.35]$, and $[0.35,1.10]$, respectively.
The magnitude is shifted by $[-6,6]$ frequency bins and mixed with the original magnitude using weights $0.35$ and $0.65$.
The sign of the middle-frequency region is reversed with probability $0.5$.

\textbf{Spectral gating.}
A $7\times7$ local average of the STFT magnitude defines a local threshold with multiplier sampled from $[0.35,0.70]$.
A sigmoid gate with slope $3$ suppresses coefficients below the threshold, followed by the frequency-dependent weighting $1-0.55f^{1.5}$.

\textbf{Mixed reconstruction.}
Spectral-smoothing resynthesis is always applied first.
Quantized-codec processing is then applied with probability $0.65$, and spectral gating is independently applied with probability $0.50$.

FACodec, SpeechTokenizer, HiFiCodec, VoiceCraft, the evaluation vocoders, and the evaluation enhancement models are not used by the Attack Simulator.
EnCodec at 6~kbps and Opus at 16~kbps are the only exact compression configurations shared between training and evaluation, while the remaining model-based reconstruction conditions are unseen during training.

\subsection{Robustness Evaluation Attacks}
We evaluate the deployed watermarking system under five categories of attacks: signal distortion, traditional compression, neural codec regeneration, vocoder resynthesis, and enhancement reconstruction.
Clean speech is reported separately.
Unless otherwise stated, processed waveforms are converted back to 24~kHz and restored to the original duration before watermark recovery.

\textbf{Signal Distortion.}
\begin{itemize}
    \item \textbf{Gaussian Noise.}
    Additive white Gaussian noise is applied at 5~dB SNR.

    \item \textbf{Low-Pass Filtering.}
    A low-pass filter with a 3-kHz cutoff is applied.

    \item \textbf{Echo.}
    A single delayed reflection is added with delay and gain sampled from $[0.1,0.5]$~s and $[0.1,0.5]$, respectively.

    \item \textbf{Resampling.}
    The 24-kHz waveform is resampled to 16~kHz and then converted back to 24~kHz.

    \item \textbf{Requantization.}
    The waveform is requantized to 8-bit precision.
\end{itemize}

\textbf{Traditional Compression.}
\begin{itemize}
    \item \textbf{MP3.}
    We perform an MP3 encode--decode round trip at 16~kbps.

    \item \textbf{Opus.}
    We perform an Opus encode--decode round trip at 16~kbps.
\end{itemize}

\textbf{Neural Codec Regeneration.}
Following the taxonomy of discrete speech tokenization~\citep{guo2025codec_survey}, we include codec-based reconstruction mechanisms with different quantization and representation designs.
For each attack, the watermarked waveform is encoded into the model's discrete representation and decoded back to speech, thereby replacing the original waveform realization through a learned quantization bottleneck and decoder.

\begin{itemize}
    \item \textbf{EnCodec}
    is a general-purpose neural audio codec based on residual vector quantization (RVQ).
    We use the released 24-kHz model at 6~kbps, where the watermarked waveform is quantized into multiple residual codebook streams and reconstructed by the codec decoder.

    \item \textbf{FACodec}
    is an attribute-factorized speech codec that uses factorized vector quantization to separate content, prosody, timbre, and acoustic details.
    We use its 4.8-kbps reconstruction setting and regenerate the watermarked speech from the complete set of factorized representations.

    \item \textbf{SpeechTokenizer (STCodec)}
    combines RVQ with semantic distillation, encouraging the first quantizer to represent linguistic content while subsequent quantizers retain complementary acoustic information.
    We use the released 4-kbps configuration and reconstruct speech from the complete RVQ stream.

    \item \textbf{HiFi-Codec}
    employs group-residual vector quantization (GRVQ), combining grouped and residual quantization to obtain high-fidelity reconstruction with a small number of codebooks.
    We pass the watermarked waveform through its released encoder--quantizer--decoder pipeline and use all codebooks of the selected pretrained configuration.

    \item \textbf{VoiceCraft}
    is not a standalone codec but a token-infilling neural codec language model built on discrete EnCodec representations.
    Its released system operates on a 16-kHz EnCodec representation with four codebooks.
    We use the codec-based regeneration path to transform the watermarked waveform into discrete codec tokens and reconstruct speech through the released VoiceCraft pipeline.
\end{itemize}

\textbf{Vocoder Resynthesis.}
Unlike codec regeneration, vocoder attacks first derive the model-compatible acoustic representation from the watermarked speech and then synthesize a new waveform from that representation.
We include substantially different waveform-generation mechanisms so that robustness is not evaluated against a single vocoder family.

\begin{itemize}
    \item \textbf{WaveGlow}
    is a normalizing-flow vocoder that models the conditional waveform distribution through invertible transformations.
    We extract the corresponding Mel spectrogram from the watermarked speech and resynthesize the waveform through the pretrained flow model.

    \item \textbf{HiFi-GAN}
    is a GAN-based vocoder whose generator reconstructs waveform periodic structure from Mel-spectrogram conditioning using multi-receptive-field blocks and adversarial training.
    The watermarked speech is converted to the model-compatible Mel representation and regenerated with the released V1 generator.

    \item \textbf{BigVGAN}
    is a large-scale GAN vocoder that introduces periodic activations and anti-aliased representations for waveform synthesis.
    We resynthesize the watermarked speech from its Mel representation using the released pretrained model.

    \item \textbf{Vocos}
    is a Fourier-based neural vocoder trained with an adversarial objective.
    Instead of directly generating time-domain samples, it predicts complex Fourier coefficients from acoustic features and reconstructs the waveform by inverse Fourier transform.
    We apply the released Mel-conditioned model to the Mel representation extracted from the watermarked speech.

    \item \textbf{DiffWave}
    is a diffusion vocoder that iteratively transforms noise into a waveform conditioned on a Mel spectrogram.
    We derive the conditioning Mel spectrogram from the watermarked speech and regenerate the waveform through the pretrained reverse-diffusion process.

    \item \textbf{PriorGrad}
    is a diffusion vocoder that replaces the standard Gaussian prior with a data-dependent adaptive prior derived from the acoustic condition.
    The watermarked speech is converted to its Mel condition and resynthesized through the released PriorGrad sampling process.

    \item \textbf{WaveFM}
    is a flow-matching vocoder that learns a conditional transport process from a Mel-informed prior to the target waveform.
    We extract the Mel representation of the watermarked speech and use the released model to generate a new waveform from this condition.
\end{itemize}

Model-specific sampling rates and acoustic frontends follow their released inference configurations, and all outputs are finally converted to 24~kHz and restored to the original duration before watermark recovery.

\textbf{Enhancement Reconstruction.}
Speech-enhancement attacks operate directly on the watermarked waveform and reconstruct a cleaner estimate according to the inductive bias of the enhancement model.
We include one complex time--frequency model and one waveform-domain model to cover substantially different enhancement mechanisms.

\begin{itemize}
    \item \textbf{FRCRN}
    is a complex-domain speech-enhancement model that combines convolutional encoding with frequency recurrence and predicts a complex ideal ratio mask.
    We convert the watermarked waveform to 16~kHz, apply the released \texttt{speech\_frcrn\_ans\_cirm\_16k} model, and reconstruct the enhanced waveform from the estimated complex mask.

    \item \textbf{Demucs}
    performs speech enhancement directly in the waveform domain using an encoder--decoder architecture with skip connections and recurrent temporal modeling.
    We pass the watermarked waveform through the released \texttt{master64} denoising model and use its enhanced waveform as the attacked signal.
\end{itemize}

The enhanced outputs are converted to mono when necessary, resampled to 24~kHz, and restored to the original duration before watermark recovery.

\section{Additional Experimental Details}
\label{app:experimental_details}

\subsection{Datasets}
\label{app:datasets}

We conduct experiments on LJSpeech~\citep{Ito2017ljspeech}, LibriTTS~\citep{zen2019libritts}, and AISHELL-3~\citep{shi2021aishell}, covering different speaker and language conditions.

\textbf{LJSpeech}
is a single-speaker English corpus containing approximately 24 hours of read speech.
We use it as the primary dataset for training and in-domain evaluation.
The fixed evaluation utterances are shared across watermarking methods and attack conditions to ensure that all comparisons are made on identical speech content.

\textbf{LibriTTS}
is a multi-speaker English corpus derived from LibriSpeech and contains substantially greater speaker variability than LJSpeech.
We use LibriTTS only for out-of-distribution evaluation to examine whether the learned watermarking mechanism transfers from the single-speaker training condition to unseen English speakers.

\textbf{AISHELL-3}
is a multi-speaker Mandarin corpus.
We use it for cross-lingual out-of-distribution evaluation, providing a substantially different linguistic and speaker condition from the English training data.

For all experiments, generated speech is converted to the common 24-kHz evaluation rate when required by the native synthesis pipeline.
The same preprocessing and evaluation protocol are used across compared methods.

\subsection{Generative Speech Models}
\label{app:generative_models}

We instantiate \textsc{Thrive} on Qwen3-TTS~\citep{hu2026qwen3} and Dots.tts~\citep{lian2026dots}.
We use the officially released checkpoints and inference pipelines for both models and retain their original generation settings, including model-specific preprocessing and synthesis configurations.
The pretrained model parameters remain frozen when training \textsc{Thrive}.

\textbf{Qwen3-TTS}
represents the \textit{discrete generation paradigm}.
Its autoregressive Talker predicts discrete acoustic codec tokens, which are subsequently converted to speech by the pretrained codec decoder.
We use the official voice-cloning pipeline and introduce \textsc{Rise} only into the codec decoding stage, without modifying the generated token sequence or the original generation procedure.
The resulting waveform is produced at 24~kHz.

\textbf{Dots.tts}
provides a complementary generation paradigm based on \textit{continuous acoustic representations}.
We follow its official synthesis pipeline and use the released pretrained model and AudioVAE decoder without modifying the native generation procedure.
\textsc{Rise} is instantiated within the frozen decoder while the upstream continuous acoustic representation is kept unchanged.
The model natively generates 48-kHz speech, which is resampled to 24~kHz for the shared watermark recovery and evaluation pipeline.

The two backbones therefore expose substantially different watermarking interfaces, with Qwen3-TTS synthesizing speech from discrete codec tokens and Dots.tts decoding continuous acoustic representations.
Using the same \textsc{Thrive} framework on both models allows us to evaluate its applicability across these two representative generation paradigms.

\begin{table*}[t]
\centering
\caption{Fidelity and watermark recovery performance on LibriTTS. Values are mean $\pm$ standard deviation except FAD, which is computed at the corpus level. TPR is measured at 1\% FPR.}
\huge
\resizebox{0.98\textwidth}{!}{%
\begin{tabular}{l*{10}{c}}
\toprule
\multirow{2}{*}{Method (bps)} & \multicolumn{4}{c}{Objective Fidelity} & \multicolumn{2}{c}{Perceptual Quality} & \multicolumn{2}{c}{FAD} & \multicolumn{2}{c}{Recovery} \\
\cmidrule(lr){2-5}\cmidrule(lr){6-7}\cmidrule(lr){8-9}\cmidrule(lr){10-11}
& STOI $\uparrow$ & PESQ $\uparrow$ & SSIM $\uparrow$ & MCD $\downarrow$ & DNSMOS $\uparrow$ & NISQA $\uparrow$ & VGGish $\downarrow$ & CLAP $\downarrow$ & ACC $\uparrow$ & TPR $\uparrow$ \\
\midrule
\multicolumn{11}{c}{LibriTTS (OOD Dataset)} \\
\midrule
\multicolumn{11}{l}{\textit{Post-hoc Watermarking}} \\
\midrule
AudioSeal (16)
& 0.999\std{0.001} & 4.393\std{0.100} & 0.979\std{0.007}
& 0.689\std{0.151} & 3.182\std{0.190} & 4.260\std{0.681}
& 0.791 & 0.009 & 1.000 & 1.000 \\
WavMark (32)
& 1.000\std{0.001} & 4.246\std{0.209} & 0.947\std{0.038}
& 1.532\std{0.481} & 3.191\std{0.192} & 4.275\std{0.710}
& 1.731 & 0.027 & 1.000 & 1.000 \\
TBWM (100)
& 0.990\std{0.005} & 3.870\std{0.252} & 0.933\std{0.015}
& 1.368\std{0.149} & 3.188\std{0.208} & 4.019\std{0.679}
& 1.167 & 0.045 & 1.000 & 1.000 \\
\midrule
\multicolumn{11}{l}{\textit{Generative Watermarking}} \\
\midrule
Groot (100)
& 0.886\std{0.052} & 2.236\std{0.414} & 0.849\std{0.025}
& 1.261\std{0.149} & 2.827\std{0.325} & 3.169\std{0.653}
& 0.649 & 0.021 & 0.995 & 1.000 \\
RIWF (16)
& 0.842\std{0.064} & 1.530\std{0.291} & 0.685\std{0.063}
& 6.329\std{1.386} & 2.364\std{0.524} & 2.159\std{0.568}
& 6.900 & 0.180 & 0.938 & 0.926 \\
HFGw (20)
& 0.901\std{0.033} & 2.005\std{0.275} & 0.904\std{0.019}
& 1.321\std{0.117} & 2.274\std{0.405} & 2.214\std{0.594}
& 2.183 & 0.260 & 0.980 & 1.000 \\
\midrule
\rowcolor[HTML]{f7fcfd}
Ours [Qwen3-TTS] (16)
& 0.969\std{0.050} & 3.747\std{0.566} & 0.858\std{0.075}
& 1.781\std{0.157} & 2.271\std{0.431} & 3.159\std{1.007}
& 1.539 & 0.089 & 0.999 & 1.000 \\
\rowcolor[HTML]{f7fcfd}
Ours [Dots.tts] (16)
& 0.955\std{0.197} & 4.045\std{0.479} & 0.980\std{0.004}
& 0.739\std{0.135} & 2.617\std{0.286} & 3.827\std{0.812}
& 0.412 & 0.052 & 0.999 & 0.999 \\
\bottomrule
\end{tabular}%
}
\label{tab:fidelity_libritts}
\end{table*}

\begin{table*}[t]
\centering
\caption{Fidelity and watermark recovery performance on AISHELL-3. Values are mean $\pm$ standard deviation except FAD, which is computed at the corpus level. TPR is measured at 1\% FPR.}
\huge
\resizebox{0.98\textwidth}{!}{%
\begin{tabular}{l*{10}{c}}
\toprule
\multirow{2}{*}{Method (bps)} & \multicolumn{4}{c}{Objective Fidelity} & \multicolumn{2}{c}{Perceptual Quality} & \multicolumn{2}{c}{FAD} & \multicolumn{2}{c}{Recovery} \\
\cmidrule(lr){2-5}\cmidrule(lr){6-7}\cmidrule(lr){8-9}\cmidrule(lr){10-11}
& STOI $\uparrow$ & PESQ $\uparrow$ & SSIM $\uparrow$ & MCD $\downarrow$ & DNSMOS $\uparrow$ & NISQA $\uparrow$ & VGGish $\downarrow$ & CLAP $\downarrow$ & ACC $\uparrow$ & TPR $\uparrow$ \\
\midrule
\multicolumn{11}{c}{AISHELL-3 (OOD Dataset)} \\
\midrule
\multicolumn{11}{l}{\textit{Post-hoc Watermarking}} \\
\midrule
AudioSeal (16)
& 0.996\std{0.005} & 4.059\std{0.275} & 0.905\std{0.048}
& 2.301\std{0.907} & 2.974\std{0.359} & 3.231\std{0.787}
& 1.325 & 0.095 & 1.000 & 1.000 \\
WavMark (32)
& 0.994\std{0.011} & 4.051\std{0.381} & 0.814\std{0.112}
& 2.795\std{0.936} & 2.909\std{0.444} & 3.159\std{0.792}
& 0.993 & 0.126 & 1.000 & 1.000 \\
TBWM (100)
& 0.961\std{0.020} & 3.316\std{0.487} & 0.844\std{0.086}
& 2.287\std{0.502} & 2.964\std{0.318} & 3.193\std{0.772}
& 1.076 & 0.085 & 1.000 & 1.000 \\
\midrule
\multicolumn{11}{l}{\textit{Generative Watermarking}} \\
\midrule
Groot (100)
& 0.919\std{0.032} & 3.025\std{0.334} & 0.806\std{0.055}
& 1.161\std{0.122} & 2.958\std{0.285} & 2.892\std{0.840}
& 0.309 & 0.015 & 0.954 & 1.000 \\
RIWF (16)
& 0.873\std{0.081} & 1.793\std{0.361} & 0.637\std{0.084}
& 8.523\std{1.628} & 2.550\std{0.731} & 2.359\std{0.650}
& 2.495 & 0.253 & 0.921 & 0.873 \\
HFGw (20)
& 0.898\std{0.032} & 2.082\std{0.310} & 0.876\std{0.042}
& 1.254\std{0.159} & 2.621\std{0.314} & 2.101\std{0.715}
& 1.989 & 0.178 & 0.857 & 0.851 \\
\midrule
\rowcolor[HTML]{f7fcfd}
Ours [Qwen3-TTS] (16)
& 0.902\std{0.268} & 3.820\std{0.439} & 0.874\std{0.048}
& 1.763\std{0.146} & 2.667\std{0.353} & 3.291\std{0.798}
& 1.243 & 0.083 & 0.999 & 1.000 \\
\rowcolor[HTML]{f7fcfd}
Ours [Dots.tts] (16)
& 0.887\std{0.314} & 4.069\std{0.471} & 0.979\std{0.004}
& 0.764\std{0.138} & 2.600\std{0.325} & 3.399\std{1.045}
& 0.551 & 0.067 & 0.999 & 1.000 \\
\bottomrule
\end{tabular}%
}
\label{tab:fidelity_aishell}
\end{table*}

\begin{table*}[t]
\centering
\caption{Provenance verification under common distortions and reconstruction attacks on LibriTTS. Entries are ACC / TPR@1\%FPR and Ours-Q/D denote Qwen3-TTS/Dots.tts.}
\label{tab:robustness_libritts}
\LARGE
\setlength{\tabcolsep}{10pt}

\resizebox{0.95\textwidth}{!}{
\begin{tabular}{lcccccc>{\columncolor[HTML]{f7fcfd}}c>{\columncolor[HTML]{f7fcfd}}c}
\toprule
& \multicolumn{3}{c}{\textit{Post-hoc Watermarking}}
& \multicolumn{5}{c}{\textit{Generative Watermarking}} \\
\cmidrule(lr){2-4}
\cmidrule(lr){5-9}

& AudioSeal (16) & WavMark (32) & TBWM (100) & Groot (100) & RIWF (16) & HFGw (20) 
& \cellcolor{white}\textbf{Ours-Q (16)} & \cellcolor{white}\textbf{Ours-D (16)} \\
\midrule

Clean & \metric{1.000}{1.000} & \metric{1.000}{1.000} & \metric{1.000}{1.000} & \metric{0.995}{1.000} & \metric{0.938}{0.926} & \metric{0.980}{1.000} & \metric{0.999}{1.000} & \metric{0.999}{1.000} \\

\midrule
\noalign{\vskip -2pt}
\multicolumn{9}{l}{\textit{Signal Distortion}} \\
\noalign{\vskip -2pt}
\midrule

Gau. Noi. 
& \metric{0.853}{0.576} & \metric{0.581}{0.037} & \metric{0.707}{0.962} & \metric{\uline{0.997}}{1.000} & \metric{0.793}{0.327} & \metric{0.901}{0.958} & \metric{0.973}{0.992} & \metric{\textbf{0.998}}{1.000} \\
Low-pass 
& \metric{0.986}{0.999} & \metric{1.000}{1.000} & \metric{0.993}{1.000} & \metric{0.995}{1.000} & \metric{0.748}{0.191} & \metric{0.908}{0.968} & \metric{\textbf{0.999}}{1.000} & \metric{\uline{0.999}}{1.000} \\
Echo 
& \metric{0.975}{0.993} & \metric{0.885}{0.992} & \metric{0.956}{1.000} & \metric{\textbf{0.989}}{1.000} & \metric{0.823}{0.441} & \metric{0.957}{0.999} & \metric{\uline{0.969}}{0.987} & \metric{0.912}{0.841} \\
Resample 
& \metric{1.000}{1.000} & \metric{1.000}{1.000} & \metric{0.904}{1.000} & \metric{\uline{0.995}}{1.000} & \metric{0.733}{0.158} & \metric{0.864}{0.873} & \metric{\textbf{0.999}}{1.000} & \metric{\textbf{0.999}}{1.000} \\
Quant. 
& \metric{0.996}{1.000} & \metric{0.955}{1.000} & \metric{0.908}{1.000} & \metric{0.995}{1.000} & \metric{0.919}{0.864} & \metric{0.964}{0.999} & \metric{\textbf{0.999}}{1.000} & \metric{\uline{0.999}}{1.000} \\

\midrule
\noalign{\vskip -2pt}
\multicolumn{9}{l}{\textit{Traditional Compression}} \\
\noalign{\vskip -2pt}
\midrule

MP3 
& \metric{0.997}{1.000} & \metric{0.778}{0.733} & \metric{0.794}{1.000} & \metric{\uline{0.995}}{1.000} & \metric{0.738}{0.168} & \metric{0.878}{0.912} & \metric{\textbf{0.999}}{1.000} & \metric{\textbf{0.999}}{1.000} \\
Opus 
& \metric{0.463}{0.001} & \metric{0.968}{1.000} & \metric{0.874}{1.000} & \metric{0.755}{0.998} & \metric{0.543}{0.006} & \metric{\uline{0.879}}{0.914} & \metric{\textbf{0.999}}{1.000} & \metric{\textbf{0.999}}{1.000} \\

\midrule
\noalign{\vskip -2pt}
\multicolumn{9}{l}{\textit{Neural Codec Regeneration}} \\
\noalign{\vskip -2pt}
\midrule

EnCodec 
& \metric{0.512}{0.003} & \metric{0.487}{0.002} & \metric{0.593}{0.260} & \metric{\textbf{0.995}}{1.000} & \metric{0.791}{0.320} & \metric{0.633}{0.090} & \metric{0.891}{0.748} & \metric{\uline{0.960}}{0.976} \\
FACodec 
& \metric{0.514}{0.003} & \metric{0.504}{0.004} & \metric{0.549}{0.063} & \metric{0.688}{0.913} & \metric{0.616}{0.025} & \metric{0.574}{0.031} & \metric{\uline{0.798}}{0.345} & \metric{\textbf{0.821}}{0.433} \\
STCodec 
& \metric{0.494}{0.002} & \metric{0.488}{0.002} & \metric{0.567}{0.120} & \metric{\textbf{0.970}}{1.000} & \metric{0.670}{0.063} & \metric{0.612}{0.063} & \metric{0.778}{0.277} & \metric{\uline{0.785}}{0.299} \\
HiFiCodec 
& \metric{0.513}{0.003} & \metric{0.491}{0.003} & \metric{0.559}{0.093} & \metric{0.704}{0.957} & \metric{0.591}{0.015} & \metric{0.584}{0.037} & \metric{\uline{0.871}}{0.660} & \metric{\textbf{0.927}}{0.895} \\
VoiceCraft 
& \metric{0.499}{0.002} & \metric{0.495}{0.003} & \metric{0.555}{0.080} & \metric{\textbf{0.999}}{1.000} & \metric{0.601}{0.019} & \metric{0.533}{0.013} & \metric{0.823}{0.441} & \metric{\uline{0.826}}{0.456} \\

\midrule
\noalign{\vskip -2pt}
\multicolumn{9}{l}{\textit{Vocoder Resynthesis}} \\
\noalign{\vskip -2pt}
\midrule

WaveGlow 
& \metric{0.517}{0.003} & \metric{0.505}{0.004} & \metric{0.762}{0.999} & \metric{0.757}{0.998} & \metric{0.559}{0.008} & \metric{0.495}{0.005} & \metric{\textbf{0.944}}{0.942} & \metric{\uline{0.877}}{0.689} \\
HiFi-GAN 
& \metric{0.511}{0.003} & \metric{0.502}{0.004} & \metric{0.789}{1.000} & \metric{0.713}{0.972} & \metric{0.570}{0.010} & \metric{0.613}{0.064} & \metric{\textbf{0.869}}{0.648} & \metric{\uline{0.741}}{0.175} \\
BigVGAN 
& \metric{0.525}{0.004} & \metric{0.505}{0.004} & \metric{0.885}{1.000} & \metric{0.703}{0.954} & \metric{0.620}{0.027} & \metric{0.508}{0.007} & \metric{\textbf{0.899}}{0.784} & \metric{\uline{0.863}}{0.619} \\
Vocos 
& \metric{0.517}{0.003} & \metric{0.545}{0.014} & \metric{0.942}{1.000} & \metric{0.763}{0.999} & \metric{0.605}{0.020} & \metric{0.634}{0.092} & \metric{\textbf{0.897}}{0.776} & \metric{\uline{0.859}}{0.605} \\
DiffWave 
& \metric{0.513}{0.003} & \metric{0.484}{0.002} & \metric{0.660}{0.775} & \metric{0.732}{0.991} & \metric{0.532}{0.004} & \metric{0.502}{0.006} & \metric{\textbf{0.927}}{0.895} & \metric{\uline{0.858}}{0.596} \\
PriorGrad 
& \metric{0.504}{0.002} & \metric{0.489}{0.002} & \metric{0.662}{0.785} & \metric{0.726}{0.986} & \metric{0.527}{0.004} & \metric{0.510}{0.007} & \metric{\textbf{0.927}}{0.895} & \metric{\uline{0.858}}{0.596} \\
WaveFM 
& \metric{0.524}{0.004} & \metric{0.460}{0.001} & \metric{0.939}{1.000} & \metric{0.705}{0.959} & \metric{0.708}{0.112} & \metric{0.647}{0.112} & \metric{\uline{0.912}}{0.841} & \metric{\textbf{0.961}}{0.977} \\

\midrule
\noalign{\vskip -2pt}
\multicolumn{9}{l}{\textit{Enhancement Reconstruction}} \\
\noalign{\vskip -2pt}
\midrule

FRCRN 
& \metric{0.999}{1.000} & \metric{0.998}{1.000} & \metric{0.989}{1.000} & \metric{\textbf{0.995}}{1.000} & \metric{0.578}{0.012} & \metric{\uline{0.947}}{0.997} & \metric{0.899}{0.784} & \metric{0.852}{0.570} \\
Demucs 
& \metric{1.000}{1.000} & \metric{1.000}{1.000} & \metric{0.995}{1.000} & \metric{\textbf{0.995}}{1.000} & \metric{0.578}{0.012} & \metric{0.941}{0.995} & \metric{\uline{0.970}}{0.989} & \metric{0.887}{0.734} \\

\bottomrule
\end{tabular}
}
\end{table*}

\begin{table*}[t]
\centering
\caption{Provenance verification under common distortions and reconstruction attacks on AISHELL-3. Entries are ACC / TPR@1\%FPR and Ours-Q/D denote Qwen3-TTS/Dots.tts.}
\label{tab:robustness_aishell}
\LARGE
\setlength{\tabcolsep}{10pt}

\resizebox{0.95\textwidth}{!}{
\begin{tabular}{lcccccc>{\columncolor[HTML]{f7fcfd}}c>{\columncolor[HTML]{f7fcfd}}c}
\toprule
& \multicolumn{3}{c}{\textit{Post-hoc Watermarking}}
& \multicolumn{5}{c}{\textit{Generative Watermarking}} \\
\cmidrule(lr){2-4}
\cmidrule(lr){5-9}

& AudioSeal (16) & WavMark (32) & TBWM (100) & Groot (100) & RIWF (16) & HFGw (20) 
& \cellcolor{white}\textbf{Ours-Q (16)} & \cellcolor{white}\textbf{Ours-D (16)} \\
\midrule

Clean & \metric{1.000}{1.000} & \metric{1.000}{1.000} & \metric{1.000}{1.000} & \metric{0.954}{1.000} & \metric{0.921}{0.873} & \metric{0.857}{0.851} & \metric{0.999}{1.000} & \metric{0.999}{1.000} \\

\midrule
\noalign{\vskip -2pt}
\multicolumn{9}{l}{\textit{Signal Distortion}} \\
\noalign{\vskip -2pt}
\midrule

Gau. Noi. 
& \metric{0.859}{0.602} & \metric{0.582}{0.037} & \metric{0.670}{0.830} & \metric{0.957}{1.000} & \metric{0.860}{0.607} & \metric{0.814}{0.690} & \metric{\uline{0.974}}{0.992} & \metric{\textbf{0.998}}{1.000} \\
Low-pass 
& \metric{0.975}{0.993} & \metric{1.000}{1.000} & \metric{0.990}{1.000} & \metric{0.962}{1.000} & \metric{0.722}{0.136} & \metric{0.795}{0.608} & \metric{\textbf{0.999}}{1.000} & \metric{\uline{0.999}}{1.000} \\
Echo 
& \metric{0.952}{0.961} & \metric{0.878}{0.988} & \metric{0.970}{1.000} & \metric{0.920}{1.000} & \metric{0.819}{0.425} & \metric{0.843}{0.803} & \metric{\textbf{0.964}}{0.981} & \metric{\uline{0.963}}{0.980} \\
Resample 
& \metric{1.000}{1.000} & \metric{1.000}{1.000} & \metric{0.914}{1.000} & \metric{0.963}{1.000} & \metric{0.707}{0.110} & \metric{0.768}{0.488} & \metric{\textbf{0.999}}{1.000} & \metric{\uline{0.999}}{1.000} \\
Quant. 
& \metric{0.965}{0.983} & \metric{0.843}{0.947} & \metric{0.775}{1.000} & \metric{0.954}{1.000} & \metric{0.938}{0.928} & \metric{0.815}{0.694} & \metric{\textbf{0.999}}{1.000} & \metric{\uline{0.999}}{1.000} \\

\midrule
\noalign{\vskip -2pt}
\multicolumn{9}{l}{\textit{Traditional Compression}} \\
\noalign{\vskip -2pt}
\midrule

MP3 
& \metric{1.000}{1.000} & \metric{0.926}{1.000} & \metric{0.834}{1.000} & \metric{0.963}{1.000} & \metric{0.713}{0.120} & \metric{0.763}{0.467} & \metric{\textbf{0.999}}{1.000} & \metric{\uline{0.998}}{1.000} \\
Opus 
& \metric{0.431}{0.000} & \metric{0.965}{1.000} & \metric{0.870}{1.000} & \metric{0.686}{0.905} & \metric{0.557}{0.008} & \metric{0.769}{0.492} & \metric{\uline{0.998}}{1.000} & \metric{\textbf{0.999}}{1.000} \\

\midrule
\noalign{\vskip -2pt}
\multicolumn{9}{l}{\textit{Neural Codec Regeneration}} \\
\noalign{\vskip -2pt}
\midrule

EnCodec 
& \metric{0.490}{0.002} & \metric{0.480}{0.002} & \metric{0.557}{0.086} & \metric{\uline{0.956}}{1.000} & \metric{0.774}{0.265} & \metric{0.616}{0.067} & \metric{0.897}{0.779} & \metric{\textbf{0.960}}{0.976} \\
FACodec 
& \metric{0.502}{0.002} & \metric{0.522}{0.007} & \metric{0.563}{0.104} & \metric{0.635}{0.587} & \metric{0.624}{0.029} & \metric{0.599}{0.050} & \metric{\uline{0.787}}{0.305} & \metric{\textbf{0.827}}{0.459} \\
STCodec 
& \metric{0.492}{0.002} & \metric{0.494}{0.003} & \metric{0.558}{0.089} & \metric{\textbf{0.914}}{1.000} & \metric{0.627}{0.030} & \metric{0.625}{0.078} & \metric{0.755}{0.210} & \metric{\uline{0.802}}{0.359} \\
HiFiCodec 
& \metric{0.505}{0.002} & \metric{0.483}{0.002} & \metric{0.568}{0.123} & \metric{0.723}{0.983} & \metric{0.562}{0.009} & \metric{0.642}{0.104} & \metric{\uline{0.841}}{0.519} & \metric{\textbf{0.947}}{0.950} \\
VoiceCraft 
& \metric{0.501}{0.002} & \metric{0.485}{0.002} & \metric{0.543}{0.049} & \metric{\textbf{0.947}}{1.000} & \metric{0.600}{0.018} & \metric{0.556}{0.021} & \metric{0.787}{0.308} & \metric{\uline{0.820}}{0.430} \\

\midrule
\noalign{\vskip -2pt}
\multicolumn{9}{l}{\textit{Vocoder Resynthesis}} \\
\noalign{\vskip -2pt}
\midrule

WaveGlow 
& \metric{0.519}{0.003} & \metric{0.515}{0.006} & \metric{0.769}{0.999} & \metric{0.745}{0.996} & \metric{0.569}{0.010} & \metric{0.500}{0.006} & \metric{\textbf{0.936}}{0.922} & \metric{\uline{0.876}}{0.680} \\
HiFi-GAN 
& \metric{0.504}{0.002} & \metric{0.512}{0.005} & \metric{0.821}{1.000} & \metric{0.726}{0.987} & \metric{0.563}{0.009} & \metric{0.625}{0.078} & \metric{\textbf{0.862}}{0.616} & \metric{\uline{0.746}}{0.186} \\
BigVGAN 
& \metric{0.534}{0.005} & \metric{0.497}{0.003} & \metric{0.899}{1.000} & \metric{0.686}{0.904} & \metric{0.608}{0.021} & \metric{0.521}{0.010} & \metric{\textbf{0.880}}{0.700} & \metric{\uline{0.868}}{0.645} \\
Vocos 
& \metric{0.499}{0.002} & \metric{0.546}{0.014} & \metric{0.961}{1.000} & \metric{0.748}{0.997} & \metric{0.599}{0.018} & \metric{0.614}{0.065} & \metric{\textbf{0.892}}{0.757} & \metric{\uline{0.861}}{0.613} \\
DiffWave 
& \metric{0.503}{0.002} & \metric{0.479}{0.002} & \metric{0.628}{0.527} & \metric{0.710}{0.967} & \metric{0.528}{0.004} & \metric{0.547}{0.017} & \metric{\textbf{0.933}}{0.913} & \metric{\uline{0.865}}{0.631} \\
PriorGrad 
& \metric{0.502}{0.002} & \metric{0.486}{0.002} & \metric{0.635}{0.584} & \metric{0.706}{0.959} & \metric{0.534}{0.005} & \metric{0.543}{0.016} & \metric{\textbf{0.933}}{0.913} & \metric{\uline{0.865}}{0.631} \\
WaveFM 
& \metric{0.541}{0.005} & \metric{0.470}{0.001} & \metric{0.943}{1.000} & \metric{0.712}{0.970} & \metric{0.689}{0.084} & \metric{0.615}{0.066} & \metric{\uline{0.916}}{0.853} & \metric{\textbf{0.961}}{0.978} \\

\midrule
\noalign{\vskip -2pt}
\multicolumn{9}{l}{\textit{Enhancement Reconstruction}} \\
\noalign{\vskip -2pt}
\midrule

FRCRN 
& \metric{0.999}{1.000} & \metric{1.000}{1.000} & \metric{0.994}{1.000} & \metric{\textbf{0.959}}{1.000} & \metric{0.554}{0.007} & \metric{0.840}{0.794} & \metric{\uline{0.896}}{0.770} & \metric{0.824}{0.446} \\
Demucs 
& \metric{1.000}{1.000} & \metric{1.000}{1.000} & \metric{0.992}{1.000} & \metric{\uline{0.959}}{1.000} & \metric{0.604}{0.020} & \metric{0.824}{0.732} & \metric{\textbf{0.981}}{0.997} & \metric{0.865}{0.631} \\

\bottomrule
\end{tabular}
}
\end{table*}

\subsection{Out-of-Distribution Generation Fidelity and Clean Recovery}
Tables~\ref{tab:fidelity_libritts} and~\ref{tab:fidelity_aishell} further evaluate out-of-distribution generalization and highlight three findings.
\ding{172} Watermark recovery remains near-perfect on both datasets. Qwen3-TTS and Dots.tts achieve 0.999 ACC on LibriTTS with TPRs of 1.000 and 0.999, respectively, and both reach 0.999 ACC and 1.000 TPR on AISHELL-3. This consistently exceeds the recovery accuracy of the generative baselines.
\ding{173} Generation fidelity also transfers well across domains. Dots.tts retains particularly strong spectral fidelity, achieving SSIM/MCD of 0.980/0.739 on LibriTTS and 0.979/0.764 on AISHELL-3, while Qwen3-TTS maintains competitive intelligibility and perceptual quality.
\ding{174} The consistent behavior across multi-speaker English and Chinese speech shows that \textsc{Thrive} is not tied to the training speaker or language, while remaining effective on both discrete-token and continuous-representation TTS backbones.

\subsection{Out-of-Distribution Provenance Verification}
Tables~\ref{tab:robustness_libritts} and~\ref{tab:robustness_aishell} further evaluate provenance verification under out-of-distribution speech, using multi-speaker English LibriTTS and cross-lingual Mandarin AISHELL-3.
Three observations emerge.

\ding{172} \textsc{Thrive} preserves its robustness to routine processing under both speaker and language shifts.
Both backbones remain near saturation under filtering, resampling, requantization, MP3, and Opus compression, with only limited variation under Gaussian noise and echo.
This indicates that the strong robustness to conventional processing observed in-domain is not tied to the LJSpeech speaker or language distribution.
\ding{173} The reconstruction-dependent robustness profiles observed in the main evaluation remain remarkably consistent out of distribution.
Across both LibriTTS and AISHELL-3, Dots.tts achieves higher ACC than Qwen3-TTS on all five neural codec regeneration attacks, whereas Qwen3-TTS is stronger on six of the seven vocoder resynthesis models and on both enhancement reconstruction models.
WaveFM remains the notable vocoder case favoring Dots.tts.
The same cross-backbone pattern appears across in-domain and OOD speech, suggesting that reconstruction robustness is governed more strongly by the interaction between the generation representation and reconstruction mechanism than by the particular speaker or language being processed.
\ding{174} Distribution shift introduces substantially less variation than changing the reconstruction mechanism itself.
Averaged over the model-driven reconstruction attacks, Qwen3-TTS and Dots.tts retain ACCs of 88.6\% and 86.3\% on LibriTTS and 87.8\% and 86.3\% on AISHELL-3, respectively, remaining close to their in-domain performance.
TPR exhibits greater attack-specific variation, particularly for several neural codecs, but its relative behavior across reconstruction families is similarly preserved.

\begin{table*}[t]
\centering
\Huge
\caption{Robustness of \textsc{Thrive} under compound attacks on LJSpeech.
Attacks are applied from left to right, and entries report ACC / TPR@1\%FPR.}
\setlength{\tabcolsep}{7pt}
\resizebox{0.95\textwidth}{!}{
\begin{tabular}{lccccc}
\toprule
& GN$\rightarrow$Echo
& LP$\rightarrow$GN
& Echo$\rightarrow$MP3
& GN$\rightarrow$EnCodec
& Opus$\rightarrow$MP3 \\
\midrule
Qwen3-TTS
& \metric{0.897}{0.776}
& \metric{0.972}{0.991}
& \metric{0.953}{0.964}
& \metric{0.903}{0.802}
& \metric{0.997}{1.000} \\

Dots.tts
& \metric{0.941}{0.934}
& \metric{0.984}{0.998}
& \metric{0.956}{0.969}
& \metric{0.928}{0.897}
& \metric{0.999}{1.000} \\

\midrule
& Opus$\rightarrow$EnCodec
& HiFiCodec$\rightarrow$DiffWave
& FACodec$\rightarrow$BigVGAN
& VoiceCraft$\rightarrow$HiFiCodec
& Vocos$\rightarrow$DiffWave \\
\midrule
Qwen3-TTS
& \metric{0.919}{0.864}
& \metric{0.791}{0.321}
& \metric{0.748}{0.193}
& \metric{0.716}{0.124}
& \metric{0.820}{0.429} \\

Dots.tts
& \metric{0.972}{0.991}
& \metric{0.801}{0.354}
& \metric{0.743}{0.179}
& \metric{0.768}{0.247}
& \metric{0.752}{0.201} \\
\bottomrule
\end{tabular}
}

\label{tab_compound_robustness}
\end{table*}

\subsection{Robustness under Compound Attacks}
To further examine robustness beyond isolated transformations, we evaluate \textsc{Thrive} under sequential compound attacks, with the two operations applied from left to right.
The evaluated conditions combine signal processing, conventional compression, neural codec processing, and sequential model-driven reconstruction.
GN denotes Gaussian noise at 15~dB SNR.
For compression and codec processing, MP3 and Opus operate at 32~kbps, EnCodec at 12~kbps, HiFiCodec at 4~kbps, and FACodec at 4.8~kbps.

Table~\ref{tab_compound_robustness} highlights three observations.
\ding{172} \textsc{Thrive} remains robust under sequential combinations of signal processing, compression, and codec operations, with both backbones maintaining strong recovery across all evaluated conditions.
\ding{173} Sequential model-driven reconstruction imposes a substantially stronger challenge.
Nevertheless, both backbones retain clearly above-chance recovery across all evaluated compositions, with ACC ranging from 0.716 to 0.820, indicating that substantial watermark information remains recoverable after multiple learned reconstruction stages.
\ding{174} The relative robustness of the two backbones continues to vary across reconstruction compositions.
Neither backbone consistently dominates, further supporting our observation that watermark robustness depends on the interaction between the generation representation and downstream reconstruction mechanism.

\subsection{Qualitative Fidelity Analysis}
\begin{figure}
    \centering
    \includegraphics[width=0.95\linewidth]{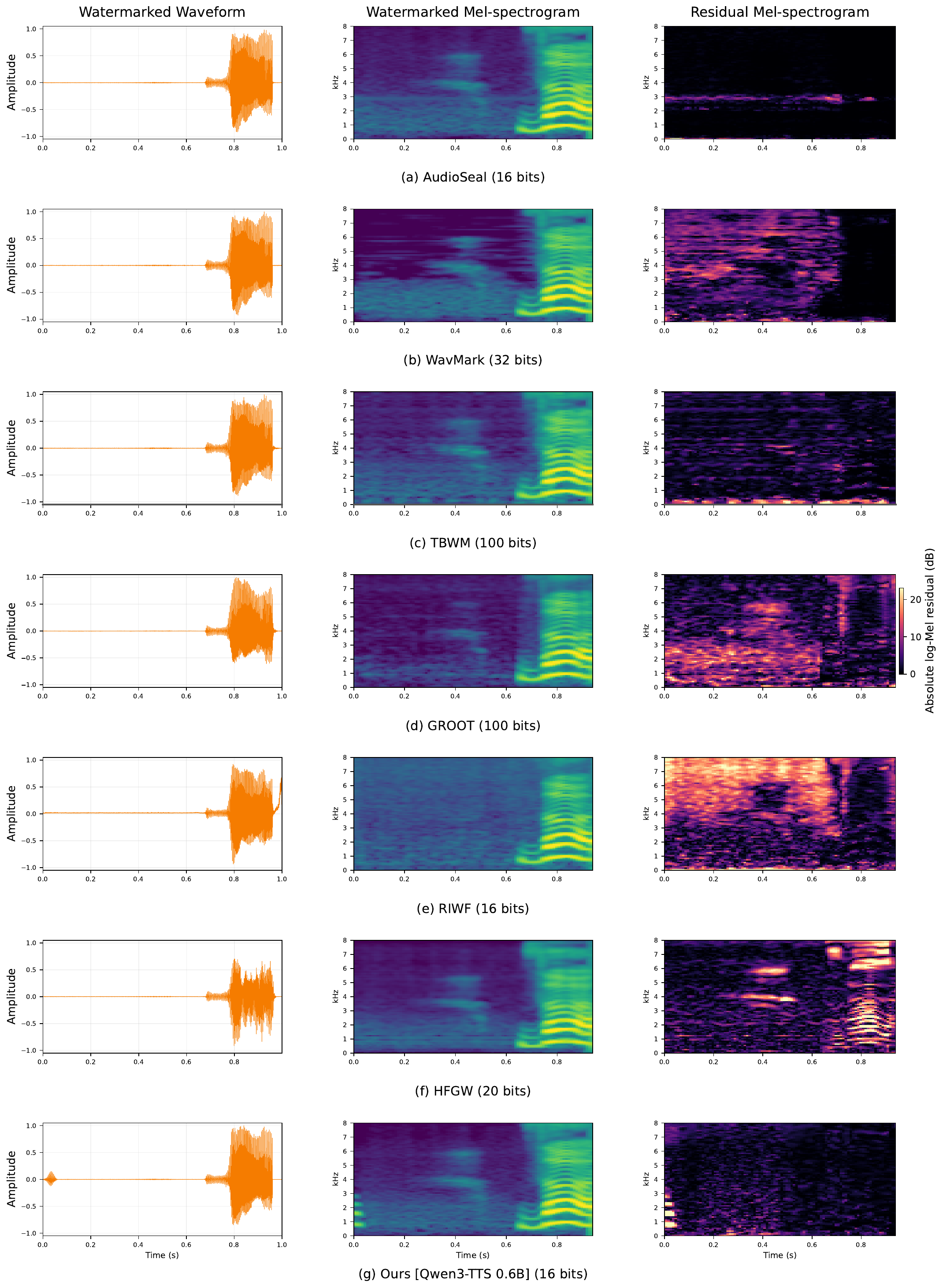}
    \caption{Qualitative visualization of fidelity on Qwen3-TTS. Each row presents the watermarked waveform, Mel-spectrogram, and residual Mel-spectrogram produced by a different watermarking method.}
    \label{fig_visual_qwen}
\end{figure}

\begin{figure}
    \centering
    \includegraphics[width=0.95\linewidth]{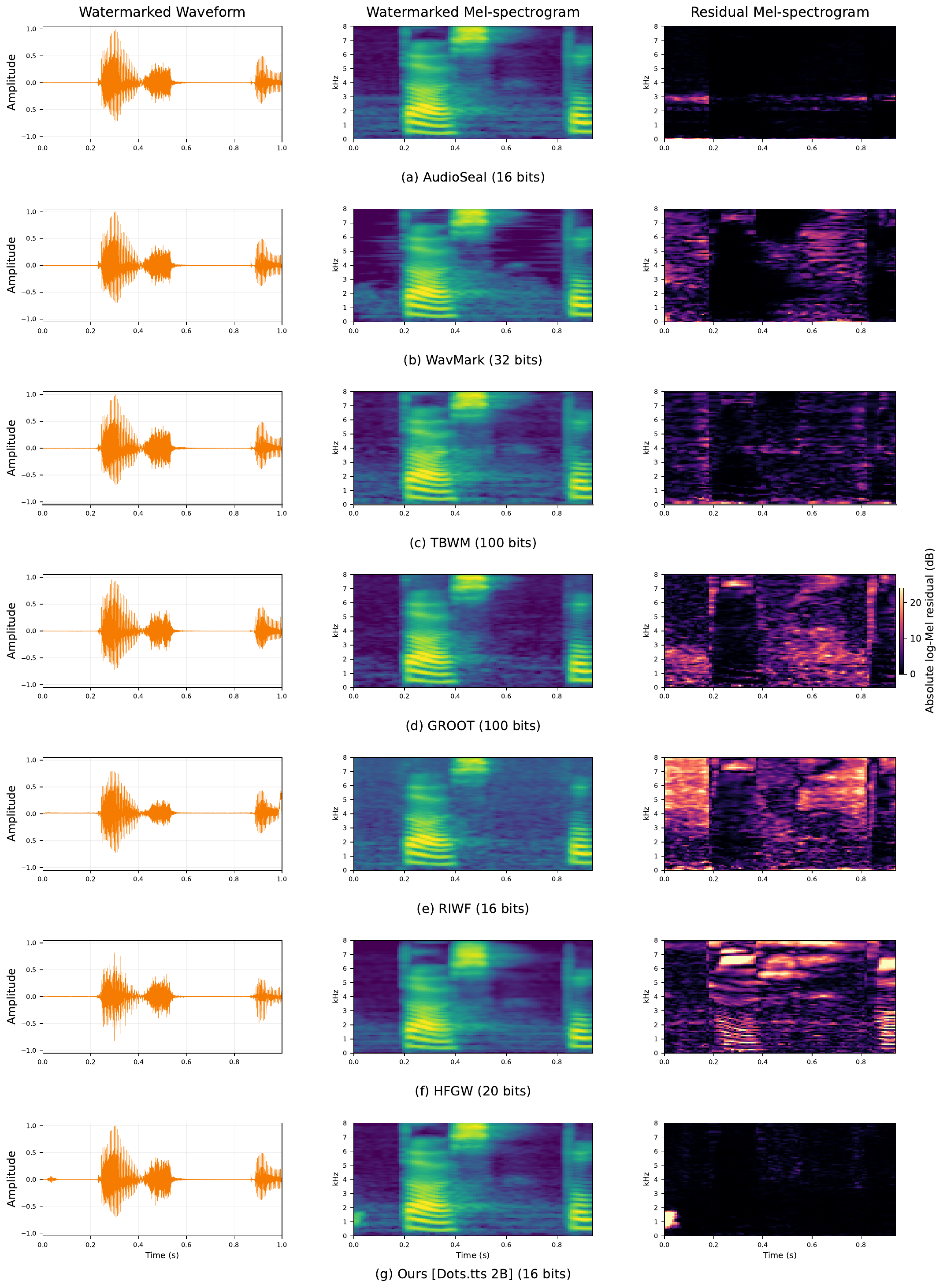}
    \caption{Qualitative visualization of fidelity on Dots.tts. Each row presents the watermarked waveform, Mel-spectrogram, and residual Mel-spectrogram produced by a different watermarking method.}
    \label{fig_visual_dots}
\end{figure}

Fig.~\ref{fig_visual_qwen} and~\ref{fig_visual_dots} provide a qualitative comparison of watermarking fidelity on Qwen3-TTS and Dots.tts.
\ding{172} Across both backbones, \textsc{Thrive} largely preserves the waveform and Mel-spectrogram structure of the generated speech, with relatively weak residual responses.
\ding{173} In comparison, several generative baselines, particularly Groot, RIWF, and HFGw, exhibit broader or stronger residual patterns across time and frequency, indicating larger acoustic deviations from the corresponding unwatermarked speech.
\ding{174} Similar fidelity patterns are observed on both Qwen3-TTS and Dots.tts, indicating that \textsc{Thrive} preserves generation quality consistently across discrete-token and continuous-representation backbones.

\section{Discussion and Limitations}
Generative speech watermarking has been developed mainly for conventional two-stage TTS pipelines consisting of an acoustic model and a vocoder.
Recent efforts have begun to extend watermarking to modern autoregressive TTS, but existing approaches focus on discrete-token generation and primarily support zero-bit detection, leaving continuous-representation autoregressive TTS unexplored.
Extending this paradigm to multi-bit watermarking is therefore non-trivial, as the watermark must be integrated into the generation process without compromising synthesis quality.
The challenge is further amplified after release, because learned reconstruction models can preserve, reorganize, or suppress watermark structures differently, making reliance on a fixed carrier or recovery representation increasingly fragile.
\textsc{Thrive} therefore extends generative watermarking to both discrete-token and continuous-representation autoregressive TTS, while supporting multi-bit identifiers in both settings.

Our primary focus is robustness to heterogeneous model-driven reconstruction, which we view as a basic requirement for practical generative speech watermarking.
While our evaluation covers diverse neural codec, vocoder, and enhancement reconstruction mechanisms, more adaptive attacks may optimize transformations specifically against the watermarking system.
Robustness against such adaptive reconstruction remains an important direction for future work.

\end{document}